\documentclass[useAMS,usenatbib]{mn2e}
\usepackage{epsfig}
\usepackage{graphicx}
\input epsf

\newcommand{\xmm}{{\it XMM~\/}}
\newcommand{\xmmn}{{\it XMM-Newton~\/}}
\newcommand{\asca}{{\it ASCA~\/}}
\newcommand{\chandra}{{\it Chandra~\/}}
\newcommand{\hst}{{\it Hubble Space Telescope~\/}}
\newcommand{\rosat}{{\it ROSAT~\/}}
\newcommand{\einstein}{{\it Einstein~\/}}

\def\ergcms{{\rm ~erg~cm^{-2}~s^{-1}}}

\def\ergsec{{\rm ~erg~s^{-1}}}

\def\H0{{\rm ~km~s^{-1}~Mpc^{-1}}}

\def\la{\mathrel{\hbox{\rlap{\hbox{\lower4pt\hbox{$\sim$}}}{\raise2pt\hbox{$<$}}}}}
\def\ga{\mathrel{\hbox{\rlap{\hbox{\lower4pt\hbox{$\sim$}}}{\raise2pt\hbox{$>$}}}}}
\def\d25{D$_{25}$}

\def\hii{H{\small II}$~$}

\def\deg{\hbox{$^\circ$~\/}}
\def\arcm{\hbox{$^\prime$~\/}}
\def\arcs{\hbox{$^{\prime\prime}$~\/}}

\title[An \xmmn view of M101: I] {An \xmmn view of M101: I. The luminous X-ray source population}
\author[L. Jenkins et al.]
	{L.P.\ Jenkins$^1$, T.P.\ Roberts$^1$, R.S.\ Warwick$^1$, R.E.\ Kilgard$^{1,2}$, M.J.\ Ward$^1$\\  
$^1$ X-ray \& Observational Astronomy Group, Dept. of Physics \& Astronomy, University of Leicester, University Road, Leicester LE1 7RH, U.K.\\
$^2$ Harvard-Smithsonian Center for Astrophysics, 60 Garden Street, Cambridge, MA 02138, USA. \\}

\pagerange{\pageref{firstpage}--\pageref{lastpage}}
\pubyear{2003}

\begin{document}

\maketitle

\label{firstpage}

\begin{abstract}

We present the first results of an \xmmn EPIC observation of the luminous X-ray source population in the face-on supergiant spiral galaxy M101. We have studied the spectral and temporal properties of the fourteen most luminous sources, all of which have intrinsic X-ray luminosities exceeding the Eddington limit for a 1.4$M_{\odot}$ neutron star, with a subset in the ultraluminous X-ray source (ULX) regime  ($L_X\geq10^{39} \ergsec$).  Eleven sources show evidence of short-term variability, and most vary by a factor of $\sim$2--4 over a baseline of 11-24 years, providing strong evidence that these sources are accreting X-ray binary (XRB) systems. Our results demonstrate that these sources are a heterogeneous population, showing a variety of spectral shapes. Interestingly, there is no apparent spectral distinction between those sources above and below the ULX luminosity threshold. Nine sources are well-fit with either simple absorbed disc blackbody or powerlaw models. However in three of the four sources best-fit with powerlaw models, we cannot exclude the disc blackbody fits and therefore conclude that, coupled with their high luminosities, eight out of nine single-component sources are possibly high state XRBs. The nuclear source (XMM-10) has the only unambiguous powerlaw spectrum ($\Gamma\sim2.3$), which may be evidence for the presence of a low-luminosity AGN (LLAGN).  The remaining five sources require at least two-component spectral fits, with an underlying hard component that can be modelled by a powerlaw continuum or, in three cases, a hot disc blackbody ($T_{in}$=0.9--1.5\,keV), plus a soft component modelled as a cool blackbody/disc blackbody/thermal plasma. We have compared the spectral shapes of nine sources covered by both this observation and an archival 100\,ks \chandra observation of M101; eight show behaviour typical of Galactic XRBs (i.e. softening with increasing luminosity), the only exception being a transient source (XMM-2) which shows little change in spectral hardness despite a factor of $\sim$30 increase in luminosity. We find no definitive spectral signatures to indicate that these sources contain neutron star primaries, and conclude that they are likely to be stellar-mass black hole XRBs (BHXBs), with black hole masses of $\sim$2--23$M_{\odot}$ if accreting at the Eddington limit. 

\end{abstract}


\section{Introduction}
\label{sec:intro}

The X-ray emission of spiral galaxies arises in a combination of discrete source populations, hot diffuse gas and an active galactic nucleus (AGN) if present (e.g. \citealt{fabbiano89}).  However, for many spiral galaxies, the overall energy output in the X-ray band is dominated by the emission from the few brightest extra-nuclear point sources \citep{roberts00}. The ``top-heavy" X-ray luminosity functions derived recently from \chandra observations of nearby spiral galaxies with heightened levels of star formation activity (e.g. \citealt{kilgard02}; \citealt{colbert03}) serves to emphasise this point.  The nature of these luminous X-ray sources, which may individually outshine the rest of the galaxy, is therefore of great interest.

The brightest of these discrete sources are the ultraluminous X-ray sources (ULXs), defined as those extra-nuclear X-ray sources possessing X-ray luminosities $>10^{39} \ergsec$, which exceeds the Eddington luminosity of a 1.4$M_{\odot}$ neutron star by a factor $\ga5$. The first examples of this class were detected by \einstein (e.g. \citealt{fabbiano87}) and they were later detected in large numbers by \rosat (\citealt{colbert99}; \citealt{roberts00}). \asca studies of ULXs demonstrated that they display ``high" state multicolour disc blackbody spectra, and in some cases spectral transitions to a ``low" powerlaw-dominated state, features in common with many Galactic black hole X-ray binary (XRB) systems.  However, the masses inferred for the compact objects in these ULXs, assuming Eddington-limited systems, are $\sim100M_{\odot}$, much larger than the masses of known Galactic black hole candidates (c.f. \citealt{mcclintock03}).  Several models have now been put forward to explain the high X-ray luminosities of ULXs. These include: sub-Eddington rate accretion onto a new class of ``intermediate-mass" (10$^2$--10$^4 M_\odot$) black holes (IMBHs, e.g. \citealt{colbert99}; \citealt{kaaret01}); XRBs with thin super-Eddington accretion disks \citep{begelman02}; XRBs emitting anisotropically \citep{king01}; and XRBs as scaled down versions of quasars (microquasars) with relativistic jets directly in our line of sight \citep{mirabel99}. A small fraction of ULXs are also known to be associated with recent supernovae (e.g. SN~1979, \citealt*{immler98}). The observational evidence does not currently exclude any of these models completely.  However it does now suggest that there are two separate populations of ULXs.  Comparatively large numbers of ULXs have been found to be associated with regions of current star formation in starburst galaxies (e.g., The Antennae (NGC~4038/39), \citealt{zezas02a}; NGC~3256, \citealt{lira02}; NGC 4485/90, \citealt{roberts02}; Arp299 (NGC~3690), \citealt*{zezas03}), and hence must be associated with the young stellar populations residing in these regions.  Other ULXs have been found in elliptical/S0 galaxies where the stellar populations are much older (e.g. \citealt*{irwin03}; \citealt{colbert03}), although it is not entirely clear whether many of these sources are associated with globular clusters within the galaxies, or are background AGN.

The detailed behavioural characteristics of XRBs in our own Galaxy may be the key to our understanding of both the nature of ULXs and other very luminous X-ray sources in nearby galaxies, particularly since at least three Galactic black-hole XRBs have been seen to reach ULX-like luminosities \citep{mcclintock03}.  However, we have previously been unable to derive sufficient detail from the study of X-ray sources in nearby galaxies for the necessary comparisons, due to the photon-limited nature of the available data.  This has changed with the large effective area spectro-imaging capabilities of {\it XMM-Newton}, where we are now able to begin probing the detailed characteristics of luminous X-ray sources in nearby galaxies outside the Local Group (e.g. \citealt{pietsch01}; \citealt{miller03a}; \citealt*{pietsch03}).

One particularly good target for such a study is the nearby supergiant spiral galaxy M101 (NGC~5457). Its face-on aspect and low line-of-sight Galactic hydrogen column density provide an ideal opportunity to study the discrete X-ray source population in a galaxy similar to the Milky Way, away from the obscuring hydrogen in our Galactic plane. M101 has previously been observed with all of the major X-ray observatories.  Its X-ray emission was first detected with the \einstein X-ray Observatory (\citealt{mccammon84}; \citealt*{trinchieri90}). \rosat studies revealed the presence of numerous discrete sources \citep*{wangetal99} as well as a substantial diffuse component \citep{snowden95}. Most recently, the superb subarcsecond imaging resolution of \chandra has revealed $>$ 100 discrete sources in the central $\sim$8 arcminutes alone (\citealt{pence01}; \citealt{mukai03}), as well as diffuse emission tracing the spiral arms described by a two-temperature thermal plasma \citep{kuntz03}. This galaxy was also thought to host five hypernova remnants (HNRs), defined as luminous discrete X-ray sources with supernova remnant (SNR) counterparts \citep{wang99}. However, the superior spatial resolution and astrometry of \chandra has now shown that two of these X-ray sources are not in fact directly coincident with the SNRs, and that another shows temporal variability indicating an accreting XRB \citep{snowden01}. The two remaining sources (MF37 \& NGC5471B) were not covered by the original \chandra observation but are within the field of view of the \xmmn observation presented in this paper.

This is the first of a series of papers on the \xmmn observation of M101. Here we present the results of the spectral and timing properties of the luminous X-ray source population, restricting ourselves to those sources with sufficient counts to perform meaningful spectral fitting. All of these sources have luminosities $\ga2\times10^{38}$ $\ergsec$, i.e. above the Eddington limit for accretion on to a 1.4$M_{\odot}$ neutron star. We also compliment this data set with an analysis of two archival \chandra observations of M101 (see section~\ref{sec:chan_obs}). The layout of this paper is as follows. In sections ~\ref{sec:obs} \& ~\ref{sec:ID}, we outline the data reduction and source selection criteria. In sections ~\ref{sec:spectra} \& ~\ref{sec:timing} we describe the spectral fitting and timing analysis techniques. This is followed in section ~\ref{sec:sourceprops} by a detailed discussion of the spectral and temporal results for each source, and in section ~\ref{sec:discuss} we discuss discuss the overall properties of this bright X-ray source population. Our conclusions are summarized in section ~\ref{sec:summary}.

\section{Observations \& Data Reduction}
\label{sec:obs}

\begin{figure*}
\centering
\includegraphics[width=17cm,height=16.5cm]{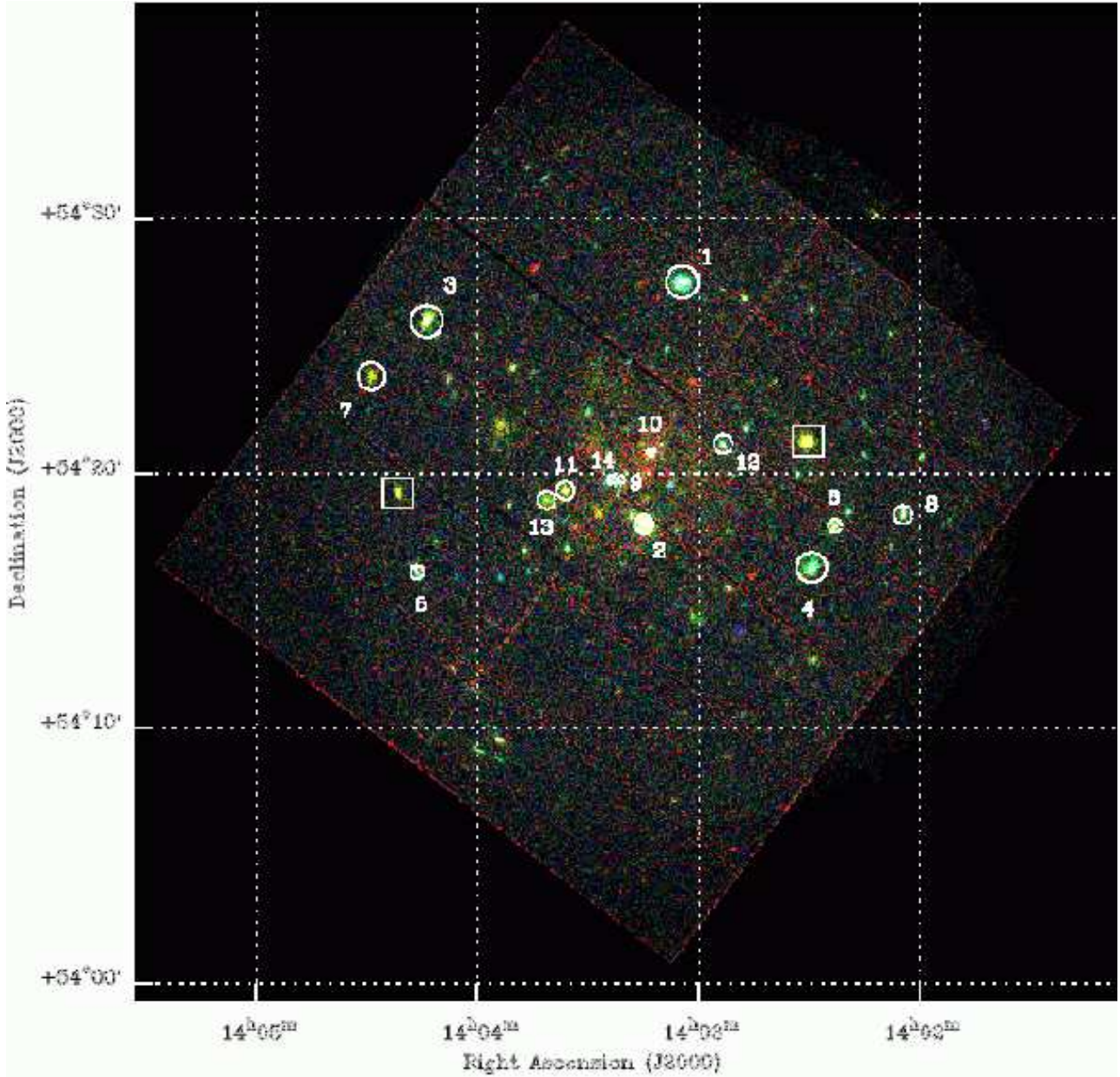}
\caption{\xmm 3-colour image of the M101 field (stacked MOS \& PN). Colours correspond to 0.3--0.5\,keV (red), 0.5--2.0\,keV (green) \& 2.0--4.5\,keV (blue). The 14 luminous X-ray sources are marked with circles denoting the spectral extraction radius used. Two bright foreground stars in the field are marked with squares.}
\label{fig:xmmim}
\end{figure*}

\subsection{The \xmmn Observation}
\label{sec:xmm_obs}

M101 was observed with \xmmn for 42.8\,ks on the 4th June 2002 (ObsID 0104260101) as part of the \xmmn Survey Science Centre (SSC) Consortium Guaranteed Time (GT) programme. The EPIC MOS-1, MOS-2 \& PN cameras were operated with medium filters in ``Prime Full Window'' mode, which utilizes the full $\sim30$ arcminute field of view of {\it XMM-Newton}. This completely encompasses the galaxy's \d25 isophotal ellipse diameter of 23.8 arminutes \citep{tully88} and corresponds to a spatial coverage of 62.8\,kpc at the distance of M101 (7.2\,Mpc, \citealt{stetson98}). The data were pipeline-processed using the {\small SAS} ({\it Science Analysis Software}) {\small v5.3.2} and all subsequent data analysis was carried out using {\small SAS v5.4.1}. Full-field light curves were accumulated for the three exposures to check for high background intervals of soft proton flares. There are numerous small flares throughout the exposure, and the four most prominent peaks (each with count rates greater than $\sim$100 counts per second in the PN) were cut from subsequent data analysis, leaving a net good time for each camera of 36.7\,ks (MOS-1 \& MOS-2) and 36.6\,ks (PN).  All data (images and spectra) were created using patterns corresponding to single \& double pixel events for the PN (0 \& 1--4), and patterns 0-12 for the MOS cameras. We also set FLAG=0 to reject events from bad pixels and events too close to the CCD chip edges.

\subsection{The \chandra Observations}
\label{sec:chan_obs}

In order to study the changing spectral and temporal properties of these sources, we compliment our \xmm analysis with recent \chandra data of M101. Two \chandra observations of M101 have been performed at the time of this analysis; a long ($\sim$100\,ks) observation on the 26th March 2000 (ObsID 934) and a shorter ($\sim$10\,ks) observation on the 29th October 2000 (ObsID 2065). Though an analysis of the longer observation is presented in \citet{pence01}, we have reanalyzed this data to utilize more up to date calibration files available in the \chandra {\small CALDB v2.11}, including new gain maps, improved plate scale, and improved PHA bin calculation. The standard data processing was done at the \chandra X-ray Center, and subsequent analyses were performed using the {\small CIAO} software ({\it \chandra Interactive Analysis of Observations}) {\small v2.2.1}. The data were screened for periods of high background leaving 98.2\,ks (ObsID 934) and 9.6\,ks (ObsID 2065) of good time.   A more detailed description of the data analysis is available in \citet{kilgard03}. In addition, whilst \citet{pence01} only include the back-illuminated ACIS-S3 chip in their analysis (which covers $\sim$11 percent of the \d25 ellipse of M101), we include a set of 5 ACIS chips for both observations; in addition to the S3 chip, we use the front-illuminated ACIS I2, I3, S2, and S4 chips. The long observation covers 51 percent of the \d25 ellipse of M101, but the combined data set gives 78 percent coverage and enables us to perform spectral and timing analyses of discrete sources in a uniform, consistent fashion.

\section{Source Identifications}
\label{sec:ID}

In this paper, we are studying the brightest point sources in M101 at the time of the \xmm observation. To define our subset, we set a lower limit of 300 counts accumulated in the PN camera (0.3--10\,keV), which we judge as the minimum requirement to perform detailed spectral and temporal analysis. To search for such sources, we ran the {\small SAS} script {\small EDETECTCHAIN} on 0.3--10\,keV cleaned PN and MOS images with a 3$\sigma$ detection significance threshold.  We detected fourteen sources with $>$300 counts in the PN data, and we name them `XMM-{\it n}', numbered as {\it n} in descending PN count rate order. The sources are listed with their coordinates and PN count rates in Table~\ref{table:sources}. The only exception to the above is XMM-14, a bright source which was not identified in the PN data by the source detection algorithm due to its close proximity to a chip gap. It was detected in the MOS data, so we have derived a count rate consistent with those of the other sources (i.e. corrected for chip gap losses) by faking a PN spectrum in {\small XSPEC} using the best fit absorbed powerlaw model from its combined spectra MOS and PN spectra (see section~\ref{sec:spectra}), the response matrix and ancillary response files and exposure time from a nearby source (XMM-9). The complete source list for this observation will be presented in Paper II ({\it in preparation}). 

Figure~\ref{fig:xmmim} shows an \xmm 3-colour image of the M101 data from the three EPIC cameras. Each source is labelled with its `XMM-{\it n}' source number, and the circles represent the spectral extraction radii used. The field of view of the combined \chandra dataset covers the positions of all of the \xmm sources except the brightest source in the \xmm observation (XMM-1). A \chandra source list was created from the merged data using the {\it wavdetect} algorithm in {\small CIAO}, and the full source list is presented in \citet{kilgard03} together with comprehensive details of the source detection techniques used. Eleven of the \xmm sources covered by these data were unresolved, while the remaining two (XMM-2 \& 10) are resolved into two discrete sources each. The official \chandra names (following the \chandra naming convention) are listed in Table~\ref{table:sources}. The astrometry of \chandra is generally good to $\sim$1 arcsecond on the aimpoint ACIS S-3 chip, but the accuracy does decrease with increasing off-axis angle. In general, the \xmm and \chandra positions agree to within 2 arcseconds. We have also cross-identified the positions of the fourteen sources with other X-ray observations of M101, i.e. \rosat HRI and PSPC \citep{wangetal99} and the original \chandra ACIS-S3 analysis by \citet{pence01}, for which the relevant ID numbers are also listed in Table~\ref{table:sources}. Note that XMM-2 is coincident with the \rosat PSPC source P17, which \citep{wangetal99} associate with HRI sources H22 \& H25. However, the \xmm position closely corresponds to the position of H25 (offset $\sim$7 arcseconds) rather than H22 (offset $\sim$33 arcseconds). 

\begin{figure*}
\centering
\includegraphics[width=13.5cm,height=12.8cm]{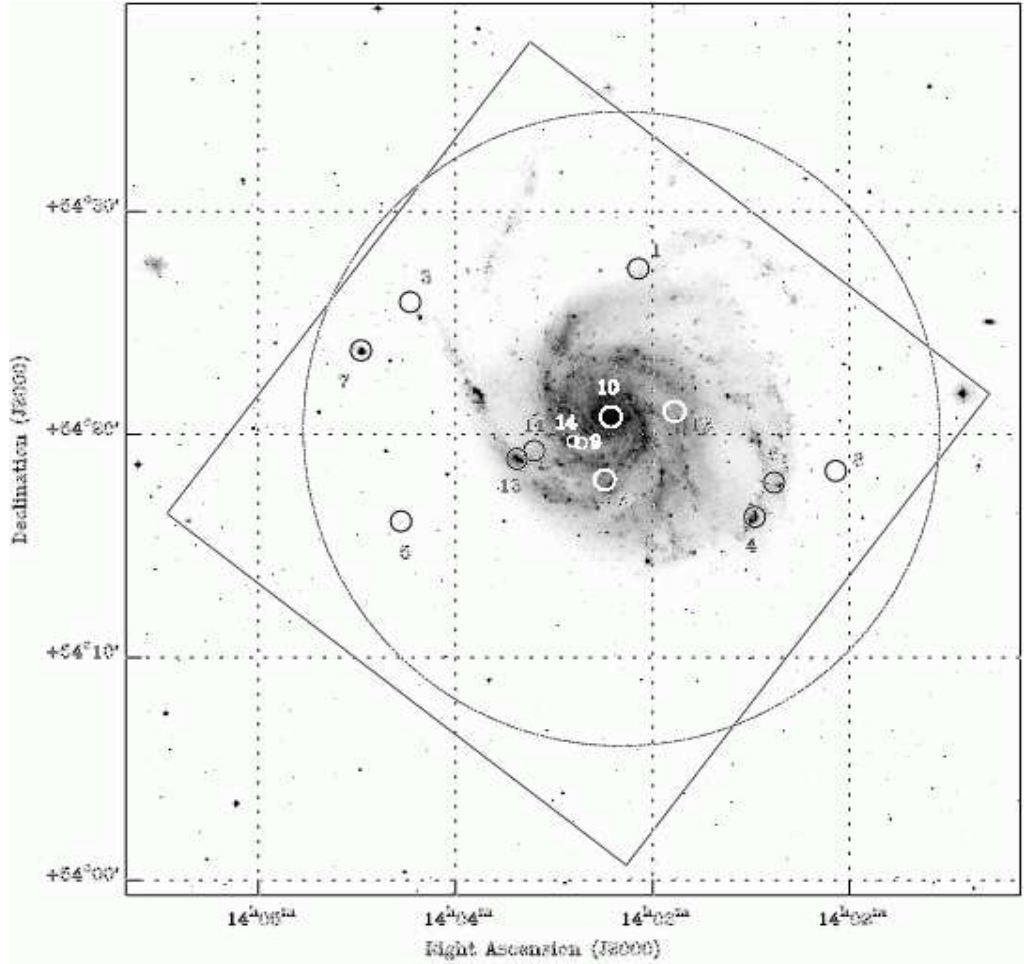}
\caption{\xmm source positions (circles) overlaid on a DSS2 optical blue image (note that source circles do not represent positional error radii and smaller circles are used to separate close sources). Large square and circular footprints denote the fields of view of the PN and MOS cameras respectively.}
\label{fig:dssim}
\end{figure*}

\begin{table*}
\caption{Source positions and cross-identifications.}
 \centering
  \begin{tabular}{lccccccc}        
\hline
Source    & RA                 & Dec                          & PN Count Rate         & Rosat ID$^a$     & \multicolumn{2}{c}{Chandra ID}      & Other ID$^{c,d}$ \\
\cline{6-7}
\\[-3mm]
          & (J2000)            & (J2000)                      & (cts ks$^{-1}$)       &                  &  Pence$^b$ & This work              &              \\
\hline
XMM-1       & $14^h03^m03.9^s$ & $+54\deg27\arcm33\arcs$ &  182.90               & P13/H19          & --          & --                    & MF37         \\ 
XMM-2       & $14^h03^m14.3^s$ & $+54\deg18\arcm05\arcs$ &  138.00               & P17/H25          & P48         & CXOU J140313.9+541811 &               \\
            &                  &                         &                       &                  & P51         & CXOU J140314.3+541807 &               \\
XMM-3       & $14^h04^m14.2^s$ & $+54\deg26\arcm03\arcs$ &  128.84               & P28/H45          & --          & CXOU J140414.3+542604 &               \\
XMM-4       & $14^h02^m28.5^s$ & $+54\deg16\arcm24\arcs$ &  75.47                & P8/H10           & --          & CXOU J140228.3+541626 & NGC 5447      \\
XMM-5       & $14^h04^m16.7^s$ & $+54\deg16\arcm13\arcs$ &  36.66                & P30/H47          & --          & CXOU J140416.8+541615 &               \\
XMM-6       & $14^h02^m22.4^s$ & $+54\deg17\arcm58\arcs$ &  32.85                & P6/H9            & --          & CXOU J140222.2+541756 &               \\
XMM-7       & $14^h04^m29.1^s$ & $+54\deg23\arcm53\arcs$ &  28.01                & P32/H49          & --          & CXOU J140429.1+542353 & NGC 5471      \\
XMM-8       & $14^h02^m03.5^s$ & $+54\deg18\arcm28\arcs$ &  27.49                & P4/H3            & --          & CXOU J140203.6+541830 &               \\
XMM-9       & $14^h03^m21.5^s$ & $+54\deg19\arcm46\arcs$ &  23.95                & P19/H29          & P70         & CXOU J140321.5+541946 &               \\
XMM-10      & $14^h03^m12.4^s$ & $+54\deg20\arcm55\arcs$ &  21.06                & P16/H23          & P38         & CXOU J140312.5+542053 &               \\
            &                  &                         &                       &                  & P40         & CXOU J140312.5+542057 &               \\
XMM-11      & $14^h03^m36.0^s$ & $+54\deg19\arcm24\arcs$ &  19.86                & P21/H36          & P104        & CXOU J140336.0+541925 & MF83          \\
XMM-12      & $14^h02^m52.8^s$ & $+54\deg21\arcm10\arcs$ &  17.60                & P11/H18          & P5          & CXOU J140252.9+542112 &               \\
XMM-13      & $14^h03^m41.2^s$ & $+54\deg19\arcm03\arcs$ &  12.11                & P22/H37          & P107        & CXOU J140341.3+541904 & NGC 5461      \\
XMM-14      & $14^h03^m24.0^s$ & $+54\deg19\arcm49\arcs$ &  10.57                & P19/H30          & P76         & CXOU J140324.2+541949 &               \\
\hline
\end{tabular}
\begin{tabular}{l}
Notes: ~$^a$\rosat ID numbers correspond to PSPC (P) \& HRI (H) detections \citep{wangetal99}. ~$^b$\chandra ID numbers from \\Pence et al. (2001). ~$^c$`MF' indicates SNR detection of \citet{matonick97}. ~$^d$`NGC' denotes giant \hii regions e.g. Williams \& \\Chu (1995).\\
\end{tabular}
\label{table:sources}
\end{table*}

Figure~\ref{fig:dssim} shows the source positions overlaid on an optical blue DSS2 image to illustrate their locations in M101. \cite{matonick97} identified 93 SNRs in M101, all of which are within the \xmm field of view. We have listed two of these (denoted MF) which we currently believe to be co-located with our X-ray sources. However, the \chandra observations \citep{snowden01} have demonstrated that two other SNRs, MF54 \& MF57, originally associated with the \rosat HRI sources H29 (XMM-9) \& H30 (XMM-14), are in fact too far offset from the SNR positions to be associated with them. Three of our sources are associated with the five giant \hii complexes in M101 (first detected in X-rays with \rosat by \citealt{williams95}), strongly suggesting that they are in some way associated with the star formation activity occurring within them.

\section{Spectral Properties}
\label{sec:spectra}

\begin{table*}
\caption{\xmmn single-component model fits.}
 \centering
  \begin{tabular*}{16cm}{lcccccccccc}        
\hline
Source  & \multicolumn{3}{c}{PL$^a$}           & & \multicolumn{3}{c}{DISKBB$^a$}            & $F_X$$^b$               & \multicolumn{2}{c}{$L_X$$^c$} \\

\cline{2-4}\cline{6-8}\cline{10-11}
\\[-3mm]
        & $N_H^d$  & $\Gamma$  & $\chi^2$/dof  & & $N_H^d$   & $T_{in}$ (keV)  & $\chi^2$/dof  & Obs                     &  Obs      & Unabs\\
\hline

XMM-1     & 2.51$^{+0.24}_{-0.23}$   & 1.90$^{+0.07}_{-0.07}$ & 305.4/291            & & 0.81$^{+0.15}_{-0.14}$  & 1.33$^{+0.07}_{-0.07}$  & {\bf{290.1/291}}   & 5.01   & 31.1 & 34.2\\
XMM-2     & 0.75$^{+0.15}_{-0.15}$   & 2.12$^{+0.09}_{-0.08}$ & {\bf{262.5/235}}$^*$ & & [$\sim$0.12]$^e$        & [$\sim$0.64]$^e$        & 600.9/235          & 3.64   & 22.6 & 28.5\\
XMM-3     & 1.75$^{+0.32}_{-0.30}$   & 2.78$^{+0.19}_{-0.17}$ & {\bf{168.1/155}}$^*$ & & $<$0.17                 & 0.58$^{+0.04}_{-0.03}$  & 185.4/155          & 2.34   & 14.5 & 31.5\\
XMM-4     & 2.04$^{+0.41}_{-0.39}$   & 2.20$^{+0.14}_{-0.06}$ & 156.5/151            & & $<$0.33                 & 1.01$^{+0.07}_{-0.07}$  & {\bf{143.6/151}}   & 2.13   & 13.2 & 13.5\\
XMM-5     & 1.47$^{+0.79}_{-0.82}$   & 2.06$^{+0.38}_{-0.27}$ & {\bf{37.0/25}}$^*$   & & $<$0.59                 & 0.98$^{+0.23}_{-0.27}$  & 48.0/25            & 1.11   & 6.9  & 9.7\\
XMM-6     & 1.86$^{+0.57}_{-0.59}$   & 2.55$^{+0.28}_{-0.24}$ & {\bf{35.0/38}}       & & $<$0.31                 & 0.72$^{+0.09}_{-0.08}$  & 40.4/38            & 0.68   & 4.2  & 8.1\\
XMM-7     & 5.72$^{+4.01}_{-2.26}$   & 7.27$^{+2.73}_{-1.76}$ & 66.8/54              & & 1.78$^{+1.61}_{-1.10}$  & 0.16$^{+0.04}_{-0.04}$  & {\bf{52.1/54}}     & 0.40   & 2.5  & 10.1\\
XMM-8     & 2.15$^{+0.86}_{-0.93}$   & 2.26$^{+0.34}_{-0.29}$ &  44.1/33             & & $<$0.79                 & 1.01$^{+0.20}_{-0.21}$  & {\bf{43.9/33}}$^*$ & 0.61   & 3.8  & 3.9\\
XMM-9     & 3.77$^{+0.94}_{-0.80}$   & 2.46$^{+0.27}_{-0.21}$ & 34.3/31              & & 1.22$^{+0.63}_{-0.52}$  & 0.96$^{+0.15}_{-0.13}$  & {\bf{30.6/31}}     & 0.57   & 3.5  & 4.2\\ 
XMM-10    & $<$0.81                  & 2.25$^{+0.41}_{-0.17}$ & {\bf{17.9/19}}       & & [$\sim$0.12]$^e$        & [$\sim$0.45]$^e$        & 39.2/19            & 0.53   & 3.3  & 3.4\\
XMM-11    & 1.65$^{+0.99}_{-0.51}$   & 3.36$^{+0.55}_{-0.35}$ & {\bf{64.0/44}}$^*$   & & $<$0.72                 & 0.34$^{+0.05}_{-0.06}$  & 69.0/44            & 0.33   & 2.1  & 6.2\\
XMM-12    & 2.69$^{+1.05}_{-1.02}$   & 1.80$^{+0.23}_{-0.25}$ & {\bf{44.1/48}}       & & 0.90$^{+0.80}_{-0.72}$  & 1.45$^{+0.49}_{-0.28}$  & 48.8/48            & 0.53   & 3.3  & 4.5\\
XMM-13    & 8.40$^{+3.06}_{-2.65}$   & $>$6.16                & 64.6/37              & & 3.17$^{+2.17}_{-1.01}$  & 0.16$^{+0.03}_{-0.04}$  & {\bf{56.3/37}}$^*$ & 0.21   & 1.3  & 10.1\\
XMM-14    & 4.18$^{+1.75}_{-1.21}$   & 1.86$^{+0.27}_{-0.23}$ & {\bf{10.0/22}}       & & 1.68$^{+1.04}_{-0.49}$  & 1.49$^{+0.35}_{-0.26}$   & 14.5/22           & 1.06   & 6.6  & 9.1\\
\hline
\end{tabular*}
\begin{tabular*}{16cm}{l}
Notes: ~$^a$Spectral models: PL=powerlaw continuum model and DISKBB=multicolour disc black body model. ~$^b${\it Observed} \\flux in the 0.3--10\,keV band, in units of $10^{-13} \ergcms$. ~$^c${\it Observed} and {\it unabsorbed} luminosities in the 0.3--10\,keV \\band, in units of $10^{38} \ergsec$ (assuming a distance to M101 of 7.2 Mpc). ~$^d$Total absorption column (including \\Galactic), in units of $10^{21}$ cm$^{-2}$. ~$^e$Parameter unconstrained due to high $\chi^2_\nu$. ~$^*$$\chi^2_\nu>1$ and therefore fits with two-\\component models are attempted (see Table.~\ref{table:spectra2}). The best-fit model for each source is highlighted in bold.\\
\end{tabular*}
\label{table:spectra1}
\end{table*}

\begin{table*}
\caption{\xmmn two-component model fits}
 \centering
  \begin{tabular*}{17cm}{llcccccccc}        
\hline
Source  & Model$^a$   & $N_H$                  & $kT/T_{in}$            &  $\Gamma$              &  $\chi^2$/dof      & $\Delta\chi^2$$^b$  & 1-P(F-test)$^c$       & \multicolumn{2}{c}{Flux Ratio$^d$} \\
\cline{9-10}
\\[-3mm]
        &             &                        & ($keV$)                &                        &                    &           &                       & Soft    & Hard      \\
\hline
XMM-2   & {\bf{PL+DISKBB}}   & {\bf{0.67$^{+0.41}_{-0.19}$}} & {\bf{0.30$^{+0.04}_{-0.06}$}} & {\bf{1.41$^{+0.25}_{-0.18}$}} &  {\bf{199.4/233}}  &  {\bf{63.1}}     & {\bf{100.0 percent}}         & {\bf{0.26}}        & {\bf{0.74}}          \\
        & PL+BBODY    & 0.38$^{+0.28}_{-0.20}$ & 0.21$^{+0.03}_{-0.03}$ & 1.51$^{+0.07}_{-0.18}$ &  200.7/233         &  61.8     & 100.0 percent         & 0.20        & 0.80  \\
        & PL+MEKAL    & 0.47$^{+0.15}_{-0.15}$ & 0.78$^{+0.11}_{-0.18}$ & 1.93$^{+0.09}_{-0.08}$ &  240.5/233         &  22.0     & $>$99.9 percent       & 0.05        & 0.95  \\
XMM-3   & {\bf{PL+MEKAL}}    & {\bf{1.37$^{+0.34}_{-0.34}$}} & {\bf{0.98$^{+0.33}_{-0.27}$}} & {\bf{2.59$^{+0.21}_{-0.19}$}} &  {\bf{160.3/153}}  &   {\bf{7.8}}     & {\bf{97.4 percent}}          & {\bf{0.05}}        & {\bf{0.95}}          \\
        & PL+BBODY    & 1.97$^{+0.51}_{-0.46}$ & 0.63$^{+0.23}_{-0.21}$ & 3.13$^{+0.36}_{-0.29}$ &  164.6/153         &   3.5     & 79.5 percent          & 0.17        & 0.83  \\
        & PL+DISKBB   & 1.94$^{+1.08}_{-0.74}$ & 0.89$^{+0.47}_{-0.34}$ & 3.22$^{+1.10}_{-0.60}$ &  165.0/153         &   3.1     & 76.1 percent          & 0.27        & 0.73  \\
XMM-5   & {\bf{PL+MEKAL}}    & {\bf{6.54$^{+1.13}_{-4.75}$}} & {\bf{0.17$^{+0.15}_{-0.05}$}} & {\bf{2.10$^{+0.47}_{-0.44}$}} &  {\bf{26.5/23}}    &  {\bf{10.5}}     & {\bf{97.8 percent}}          & {\bf{0.20}}        & {\bf{0.80}}          \\
        & PL+BBODY    & 1.91$^{+4.79}_{-0.64}$ & 0.17$^{+0.12}_{-0.08}$ & 1.43$^{+0.87}_{-1.05}$ &  30.5/23           &   6.5     & 89.1 percent          & 0.15        & 0.85  \\
        & PL+DISKBB   & 3.29$^{+3.49}_{-2.62}$ & 0.17$^{+0.30}_{-0.06}$ & 1.61$^{+0.62}_{-1.40}$ &  30.8/23           &   6.2     & 87.8 percent          & 0.16        & 0.84  \\
XMM-11  & {\bf{PL+BBODY}}    & {\bf{$<$1.19}}                & {\bf{0.19$^{+0.03}_{-0.04}$}} & {\bf{1.73$^{+1.01}_{-1.06}$}} &  {\bf{55.6/42}}    &   {\bf{8.4}}     & {\bf{94.8 percent}}          & {\bf{0.44}}        & {\bf{0.56}}          \\
        & PL+DISKBB   & 0.83$^{+0.89}_{-0.71}$ & 0.23$^{+0.09}_{-0.06}$ & 1.68$^{+1.08}_{-1.24}$ &  56.5/42           &   7.5     & 92.9 percent          & 0.49        & 0.51  \\
        & PL+MEKAL    & 0.78$^{+1.10}_{-0.59}$ & 0.99$^{+0.32}_{-0.25}$ & 2.91$^{+0.87}_{-0.32}$ &  56.8/42           &   7.2     & 91.9 percent          & 0.15        & 0.85  \\
XMM-13  & {\bf{PL+BBODY}}    & {\bf{3.41$^{+2.35}_{-1.54}$}} & {\bf{0.12$^{+0.03}_{-0.03}$}} & {\bf{0.36$^{+1.19}_{-1.56}$}} &  {\bf{34.3/35}}    &  {\bf{30.3}}     & {\bf{$>$99.9 percent}}       & {\bf{0.40}}        & {\bf{0.60}}          \\
        & PL+DISKBB   & 4.72$^{+1.83}_{-1.88}$ & 0.13$^{+0.04}_{-0.03}$ & 0.61$^{+1.02}_{-1.88}$ &  35.4/35           &  29.2     & $>$99.9 percent       & 0.41        & 0.59  \\
        & PL+MEKAL    & 5.73$^{+0.76}_{-1.81}$ & 0.18$^{+0.06}_{-0.03}$ & 0.85$^{+1.22}_{-1.33}$ &  38.4/35           &  26.2     & $>$99.9 percent       & 0.43        & 0.57  \\
       
\hline
\end{tabular*}
\begin{tabular*}{17cm}{l}
Notes: ~$^a$Spectral models and parameters as in Table~\ref{table:spectra1} except: BBODY=blackbody model and MEKAL=thermal plasma (solar \\abundances). ~$^b$Improvement in the $\chi^2$ statistic over the single component model powerlaw (PL) fit, for two extra degrees of \\freedom. ~$^c$F-test statistical probability of improvement of fit over single component model (PL). ~$^d$Fraction of total flux in the \\soft and hard (PL) model components over 0.3--10\,keV band. The best-fit model for each source is highlighted in bold.\\
\end{tabular*}
\label{table:spectra2}
\end{table*}

\begin{table*}
\caption{\chandra single component model fits.}
 \centering
  \begin{tabular*}{18.3cm}{lccccccccccc}        
\hline
Source & Chandra              & \multicolumn{3}{c}{PL$^a$}                                       & & \multicolumn{3}{c}{DISKBB$^a$}                                     & $F_X$$^b$               & \multicolumn{2}{c}{$L_X$$^c$} \\

\cline{3-5}\cline{7-9}\cline{11-12}
\\[-3mm]
       & CXOU ID              & $N_H^d$                & $\Gamma$               & $\chi^2$/dof    & & $N_H$                   & $T_{in}$ (keV)          & $\chi^2$/dof  & Obs                &  Obs & Unabs\\
\hline

XMM-1  & --                   & --                     & --                     & --              & & --                      & --                      & --            & --                      & --         & --         \\
XMM-2  & J140313.9+541811     & --                     & --                     & --              & & --                      & --                      & --            & --                      & --         & --          \\
       & J140314.3+541807     & [$\sim$1.21]$^e$       & [$\sim$2.42]$^e$       & {\bf {3/7}}     & & [$\sim$0.12]$^e$        & [$\sim$ 0.64]$^e$       & 3/7           & 0.08                    & 0.48       & 0.79       \\
XMM-3  & J140414.3+542604     & 3.51$^{+0.35}_{-0.33}$ & 3.60$^{+0.18}_{-0.16}$ & {\bf {114/120}} & & 0.57$^{+0.23}_{-0.21}$  & 0.53$^{+0.04}_{-0.04}$  & 154/120       & 2.75                    & 17.04      & 119.71     \\
XMM-4  & J140228.3+541626     & --                     & --                     & --              & & --                      & --                      & --            & --                      & --         & --         \\
XMM-5  & J140416.8+541615     & 0.88$^{+0.73}_{-0.65}$ & 1.81$^{+0.26}_{-0.23}$ & {\bf {30/58}}    & & $<$0.30               & 1.15$^{+0.25}_{-0.16}$  & 38/58         & 1.04                    & 6.43       &  7.88      \\
XMM-6  & J140222.2+541756     & --                     & --                     & --              & & --                      & --                      & --            & --                      & --         & --         \\
XMM-7  & J140429.1+542353     & --                     & --                     & --              & & --                      & --                      & --            & --                      & --         & --         \\
XMM-8  & J140203.6+541830     & --                     & --                     & --              & & --                      & --                      & --            & --                      & --         & --         \\
XMM-9  & J140321.5+541946     & 1.96$^{+0.62}_{-0.56}$ & 2.01$^{+0.27}_{-0.25}$ & 29/45           & & 0.54$^{+0.38}_{-0.34}$  & 1.07$^{+0.18}_{-0.15}$  & {\bf {25/45}} & 0.50                    & 3.09       & 3.40       \\
XMM-10 & J140312.5+542053     & $<$2.25                & 1.73$^{+0.63}_{-0.40}$ & {\bf {4/13}}    & & $<$0.61                  & 1.17$^{+0.33}_{-0.36}$  & 6/13         & 0.18                    & 1.09       & 1.29       \\
       & J140312.5+542057     & $<$3.16                & 2.07$^{+1.76}_{-0.47}$ & {\bf {13/13}}   & & $<$1.27                  & 0.53$^{+0.14}_{-0.23}$  & 19/13         & 0.13                    & 0.79       & 0.93       \\
XMM-11 & J140336.0+541925     & 0.82$^{+0.64}_{-0.53}$ & 2.66$^{+0.46}_{-0.37}$ & {\bf {24/34}}   & & $<$0.24                  & 0.44$^{+0.11}_{-0.05}$  & 42/34         & 0.30                    & 1.83       & 2.92      \\
XMM-12 & J140252.9+542112     & 0.94$^{+0.55}_{-0.47}$ & 1.49$^{+0.26}_{-0.23}$ & {\bf {30/36}}   & & [$\sim$0.15]$^e$        & [$\sim$1.61]$^e$        & 32/36         & 0.52                    & 3.24       & 3.72      \\
XMM-13 & J140341.3+541904     & 11.37$^{+3.34}_{-3.89}$& 10.0$^{+2.27}_{-2.69}$ & 33/24           & & 5.34$^{+2.09}_{-1.75}$  & 0.13$^{+0.04}_{-0.02}$  & {\bf {17/24}} & 0.12                    & 0.72       & 29.23     \\
XMM-14 & J140324.2+541949     & 4.40$^{+1.11}_{-0.95}$ & 1.72$^{+0.23}_{-0.21}$ & 32/53           & & 2.71$^{+0.62}_{-0.62}$  & 1.55$^{+0.28}_{-0.08}$  & {\bf {26/53}} & 0.87                    & 5.41       & 6.67      \\
\hline
\end{tabular*}
\begin{tabular*}{18.3cm}{l}
Notes: ~$^a$Spectral models and parameters as in Table~\ref{table:spectra1}. ~$^b${\it Observed} flux in the 0.3--8\,keV band, in units of $10^{-13} \ergcms$.\\ 
$^c${\it Observed} and {\it unabsorbed} luminosities in the 0.3--8\,keV band, in units of $10^{38} \ergsec$ (assuming a distance to M101 of 7.2 Mpc).\\
$^d$Total absorption column in units of $10^{21}$ cm$^{-2}$. ~$^e$Parameter unconstrained. The best-fit model is highlighted in bold.\\
\end{tabular*}
\label{table:spectra3}
\end{table*}

\begin{table*}
\caption{\xmmn short-term variability test results.}
 \centering
  \begin{tabular}{lccccccccc}        
\hline
Source  & Bin Size   & \multicolumn{3}{c}{Standard deviation, $\sigma$}                                        & & \multicolumn{2}{c}{$\chi^2$ statistic}  & & K-S statistic \\
\cline{3-5}\cline{7-8}\cline{10-10}
\\[-3mm]
        & (s)        &  Expected$^a$             &  Observed$^a$                         &  $P_{\sigma}$(var)  & &  $\chi^2$/dof   & $P_{\chi^2}$(var)     & & $P_{K-S}$(var)\\
\hline
XMM-1   & 200        &  2.88 $\times$ $10^{-2}$  & (3.35 $\pm$ 0.18) $\times$ $10^{-2}$  &  99.2 percent      & & 206/171         & 96.5 percent         & & -          \\
XMM-2   & 200        &  2.72 $\times$ $10^{-2}$  & (4.07 $\pm$ 0.22) $\times$ $10^{-2}$  &  $>$99.9 percent   & & 427/171         & $>$99.9 percent      & & -          \\
XMM-3   & 400        &  1.38 $\times$ $10^{-2}$  & (1.66 $\pm$ 0.13) $\times$ $10^{-2}$  &  97.6 percent      & & 104/86          & -                     & & -           \\
XMM-4   & 300        &  1.65 $\times$ $10^{-2}$  & (2.19 $\pm$ 0.14) $\times$ $10^{-2}$  &  $>$99.9 percent   & & 149/115         & 98.2 percent         & & -           \\
XMM-5   & 2000       &  2.62 $\times$ $10^{-3}$  & (3.76 $\pm$ 0.61) $\times$ $10^{-3}$  &  97.7 percent      & & 29/18           & 95.2 percent         & & -           \\
XMM-6   & 1200       &  4.33 $\times$ $10^{-3}$  & (5.14 $\pm$ 0.66) $\times$ $10^{-3}$  &  -                  & & 35/29           & -                     & & -           \\
XMM-7   & 1500       &  3.29 $\times$ $10^{-3}$  & (3.90 $\pm$ 0.56) $\times$ $10^{-3}$  &  -                  & & 12/23           & -                     & & -            \\
XMM-8   & 2000       &  2.68 $\times$ $10^{-3}$  & (4.06 $\pm$ 0.66) $\times$ $10^{-3}$  &  96.4 percent      & & 23/18           & -                     & & 97.4 percent \\
XMM-9   & 1500       &  3.50 $\times$ $10^{-3}$  & (4.22 $\pm$ 0.61) $\times$ $10^{-3}$  &  -                  & & 30/23           & -                     & & -            \\
XMM-10  & 2000       &  2.43 $\times$ $10^{-3}$  & (3.62 $\pm$ 0.59) $\times$ $10^{-3}$  &  95.8 percent      & & 34/18           & 98.7 percent         & & -            \\
XMM-11  & 1200       &  4.28 $\times$ $10^{-3}$  & (7.57 $\pm$ 0.98) $\times$ $10^{-3}$  &  99.9 percent      & & 73/29           & $>$99.9 percent      & & $>$99.9 percent\\
XMM-12  & 1500       &  3.35 $\times$ $10^{-3}$  & (5.57 $\pm$ 0.80) $\times$ $10^{-3}$  &  99.4 percent      & & 42/23           & $>$99.9 percent      & & -             \\
XMM-13  & 1500       &  3.42 $\times$ $10^{-3}$  & (4.88 $\pm$ 0.70) $\times$ $10^{-3}$  &  96.1 percent      & & 26/23           & -                     & & 99.0 percent  \\
XMM-14  & 2000       &  2.60 $\times$ $10^{-3}$  & (3.31 $\pm$ 0.54) $\times$ $10^{-3}$  &  -                  & & 22/18           & -                     & & 97.5 percent  \\
\hline
\end{tabular}
\begin{tabular}{l}
Notes: ~$^a$Expected and observed standard deviations from mean count rate (count s$^{-1}$). We only show $P$(var)$>$95 percent to highlight \\variability.\\
\end{tabular}
\label{table:var}
\end{table*}

The \xmm X-ray spectra for each source were extracted using circular regions enclosing all of the visible emission, of radii varying between 10 and 35 arcseconds. The smallest apertures were used in order to avoid contamination from nearby sources in crowded regions of the galaxy, or for sources located next to a chip gap (e.g. XMM-9, 10 \& 14). The background regions for each source were taken using as large an area as possible, close to each source without contamination from other sources and with approximately the same DET-Y distance as the source region (to ensure similar low-energy noise subtraction). We used the {\small SAS} task {\small ESPECGET} to simultaneously extract source and background spectra and to create response matrices (RMFs) and ancillary response files (ARFs) for each source. The spectra were then binned to a minimum of 20 counts per bin in order to optimise the data for $\chi^2$ statistics.

The spectral analysis of the \xmm data in this study has been performed using {\small XSPEC v11.1.0}. All errors are given at the 90 percent confidence level unless stated otherwise.  The MOS and PN spectra have been fitted simultaneously in the 0.3--10\,keV band for each source. We included a free normalisation constant to account for differences in the flux calibration of the three EPIC cameras, finding the relative normalisations to be typically within $\sim5$ percent. 

We initially fitted the spectra with simple absorbed single component spectral models: powerlaw continuum (PL), MEKAL optically-thin thermal plasma, conventional black body (BB) and a multicolour disc black body (DISKBB). This last model describes an accretion disc in its high/soft (HS) state (\citealt{mitsuda84}; \citealt{makishima86}). Each model had both a free absorption component plus an additional one fixed at the line of sight Galactic hydrogen column density (1.16$\times10^{20}$ cm$^{-2}$, \citealt{dickey90}).  In Table~\ref{table:spectra1}, we show the best fit parameters for the PL continuum and DISKBB models for each source, together with the observed fluxes\footnote{The fluxes quoted here are an average of those measured in the three EPIC cameras, except for instances where the normalisation constant of the PN differed from the MOS cameras by $>$10 percent, in which case the average MOS flux is quoted.} and observed plus unabsorbed luminosities (0.3--10\,keV) from the best fit model (highlighted in bold). We omit the BB fits from the table as they did not provide significantly better fits than the DISKBB model, except in the cases of XMM-7 and XMM-13, which we will discuss individually in sections~\ref{sec:xmm7} and ~\ref{sec:xmm13}. We also omit the MEKAL fits from the table because, in most cases, the spectra were too hard to be fitted well by this model alone; two exceptions to this are XMM-7 and XMM-14, both of which will be discussed individually in section~\ref{sec:sourceprops}.

\begin{figure*}
\centering
\parbox{6.5cm}{\scalebox{0.3}{\includegraphics{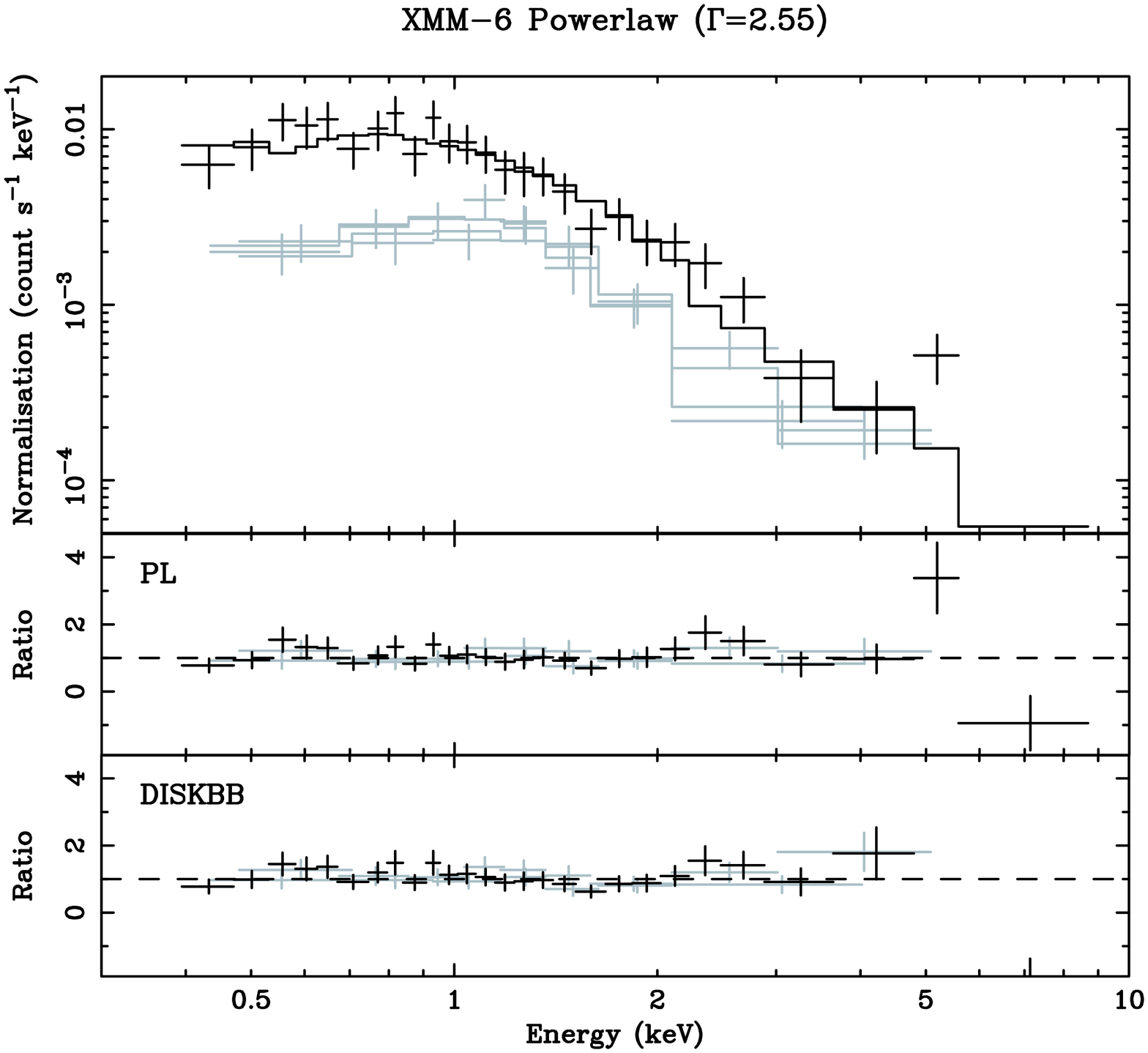}}}
\hspace*{1.5cm}
\vspace*{0.5cm}
\parbox{6.5cm}{\scalebox{0.3}{\includegraphics{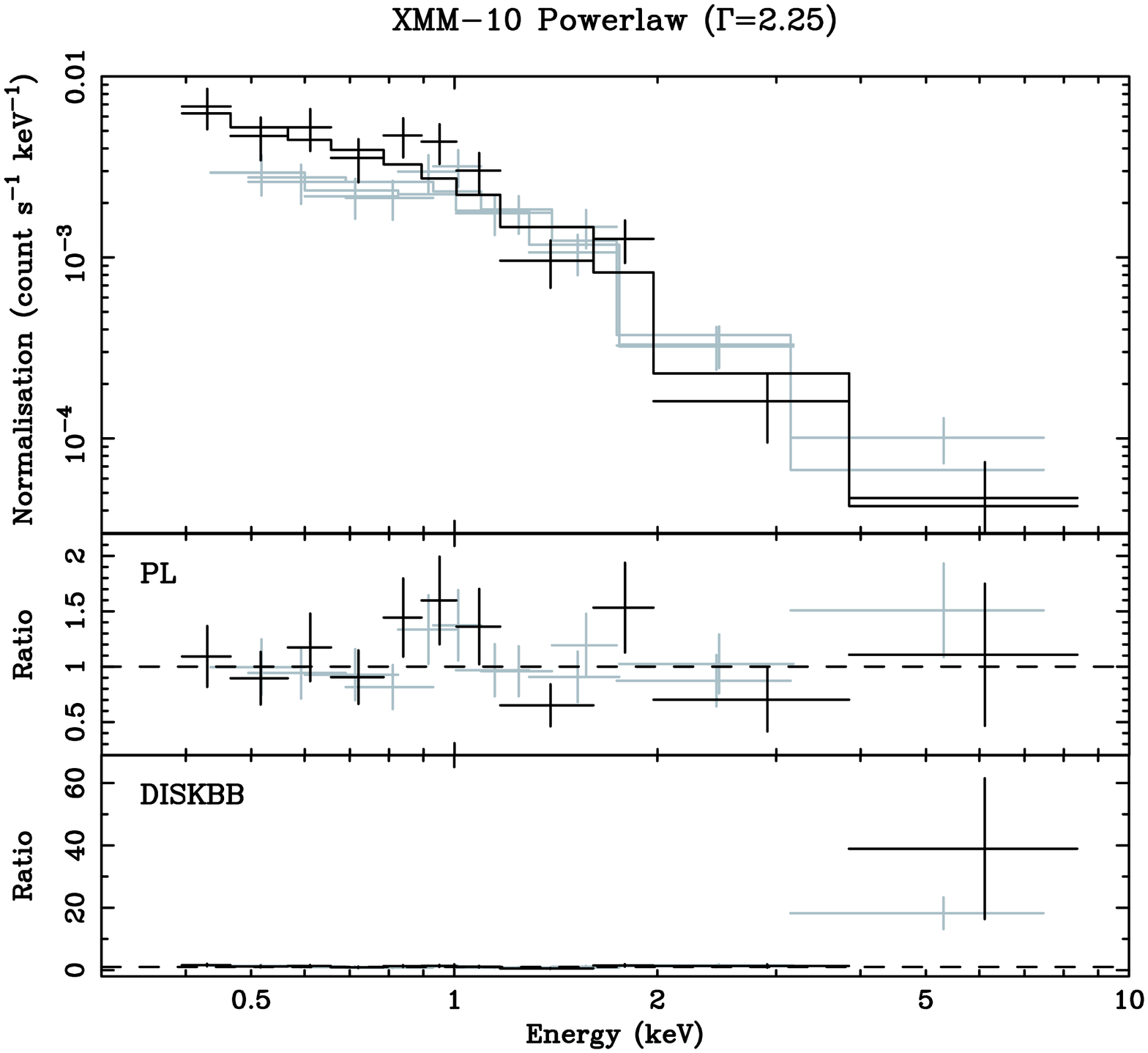}}}
\vspace*{0.5cm}
\parbox{6.5cm}{\scalebox{0.3}{\includegraphics{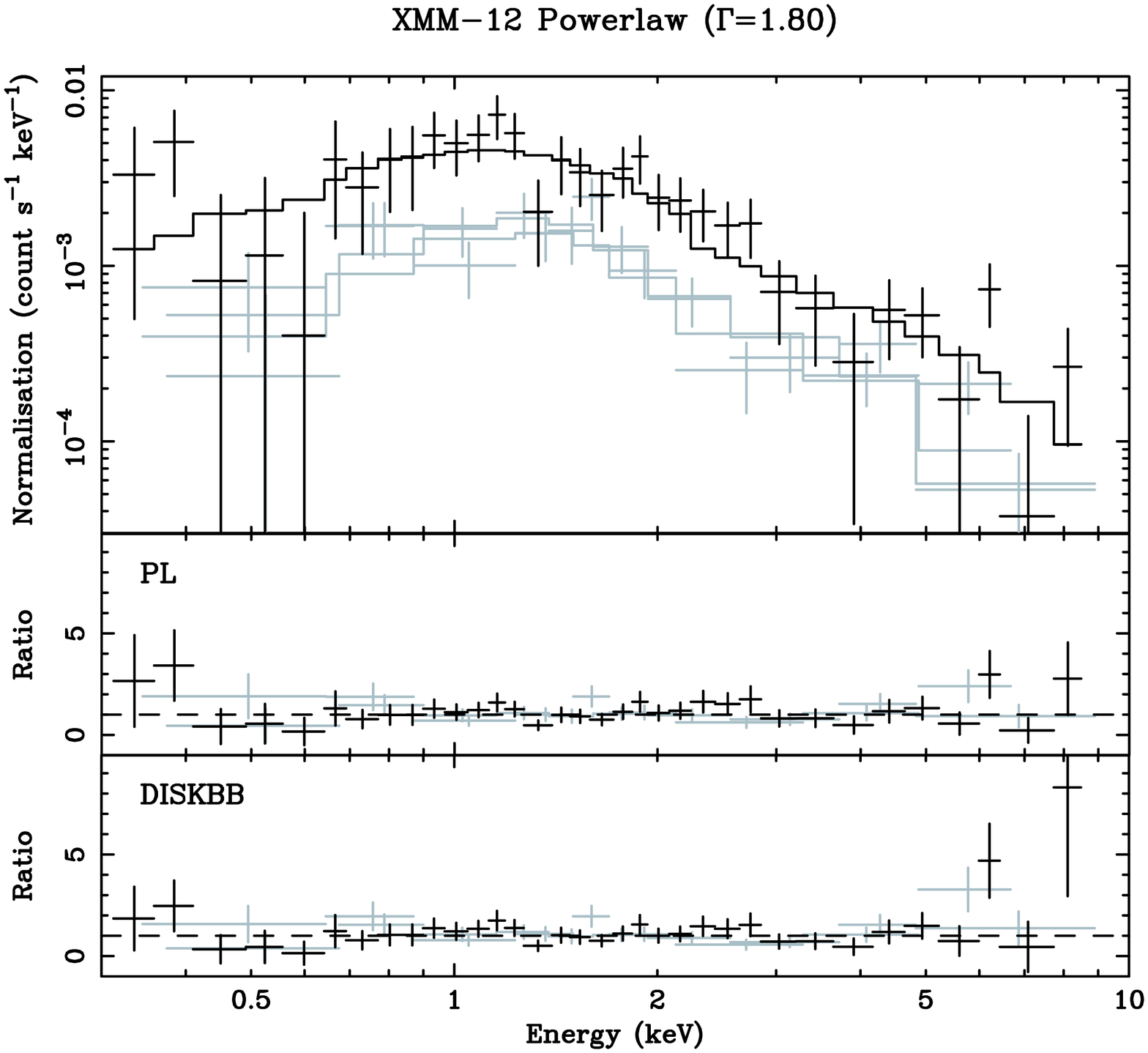}}}
\hspace*{1.5cm}
\parbox{6.5cm}{\scalebox{0.3}{\includegraphics{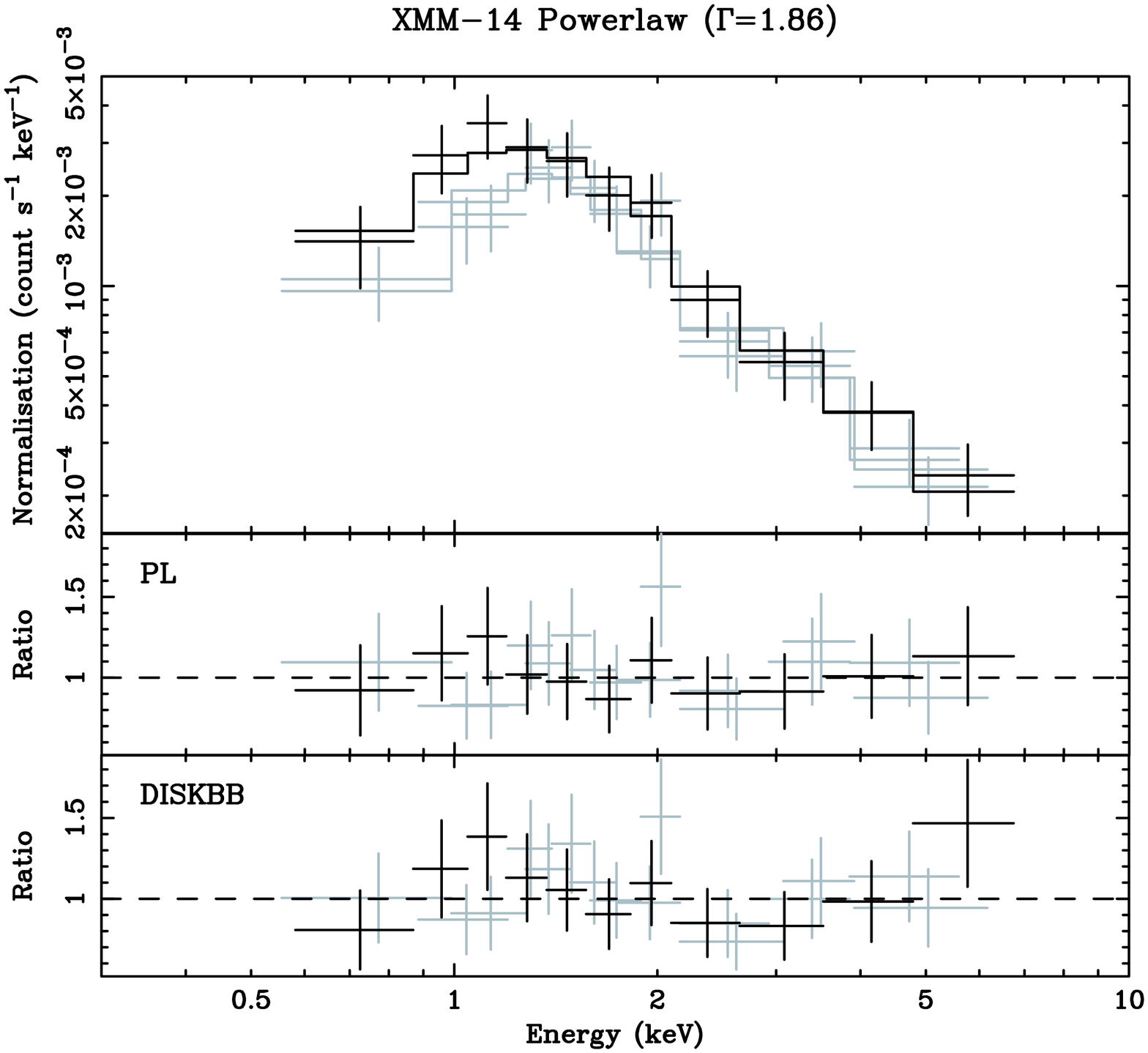}}}
\caption{\xmm spectra of the powerlaw sources. PN data points and the best fit model are shown in black; the MOS data are shown in grey. Ratios are shown for both the PL and DISKBB fits.}
\label{fig:spec1}
\end{figure*}

\begin{figure*}
\centering
\parbox{6.5cm}{\scalebox{0.3}{\includegraphics{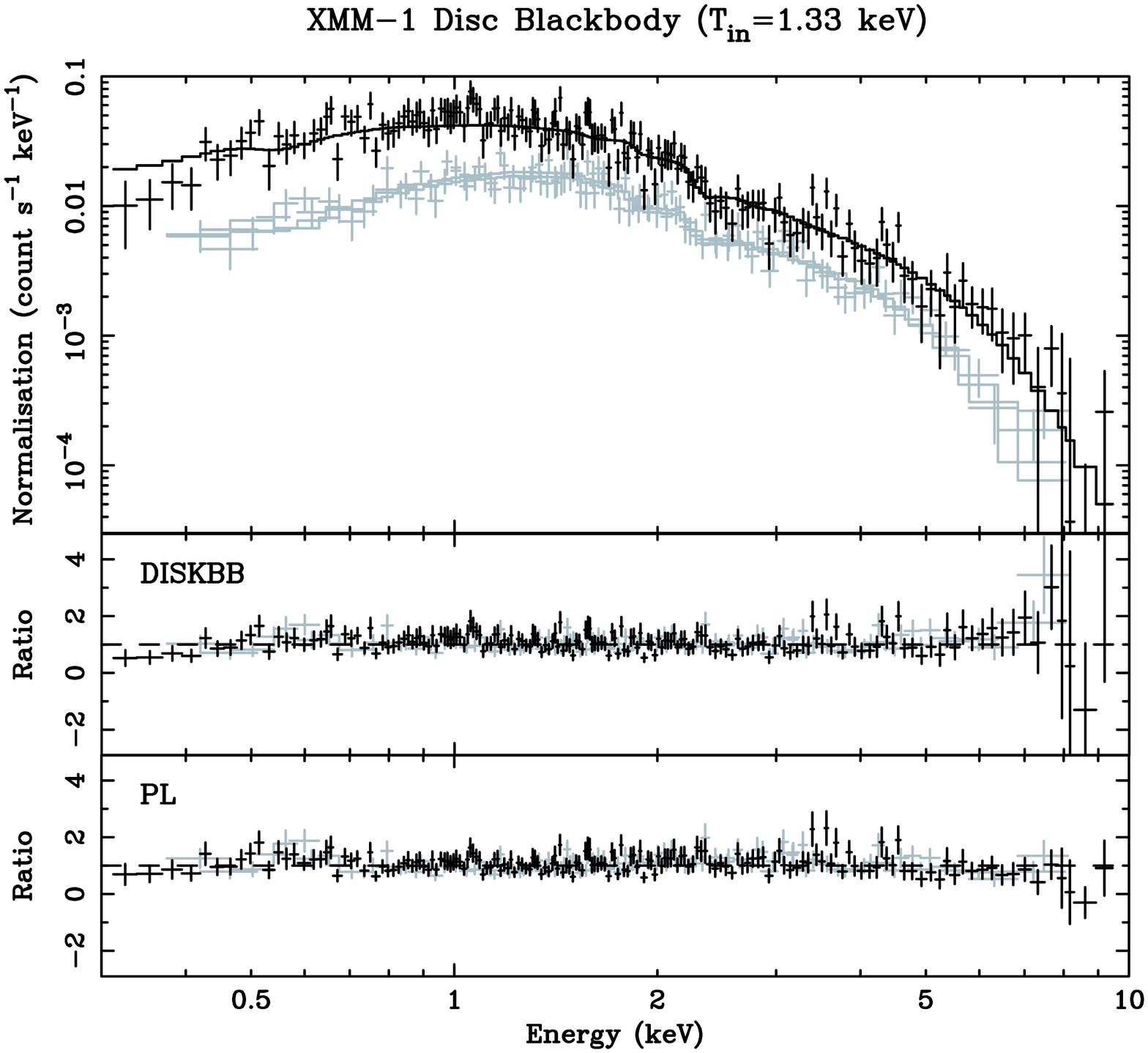}}}
\hspace*{1.5cm}
\vspace*{0.5cm}
\parbox{6.5cm}{\scalebox{0.3}{\includegraphics{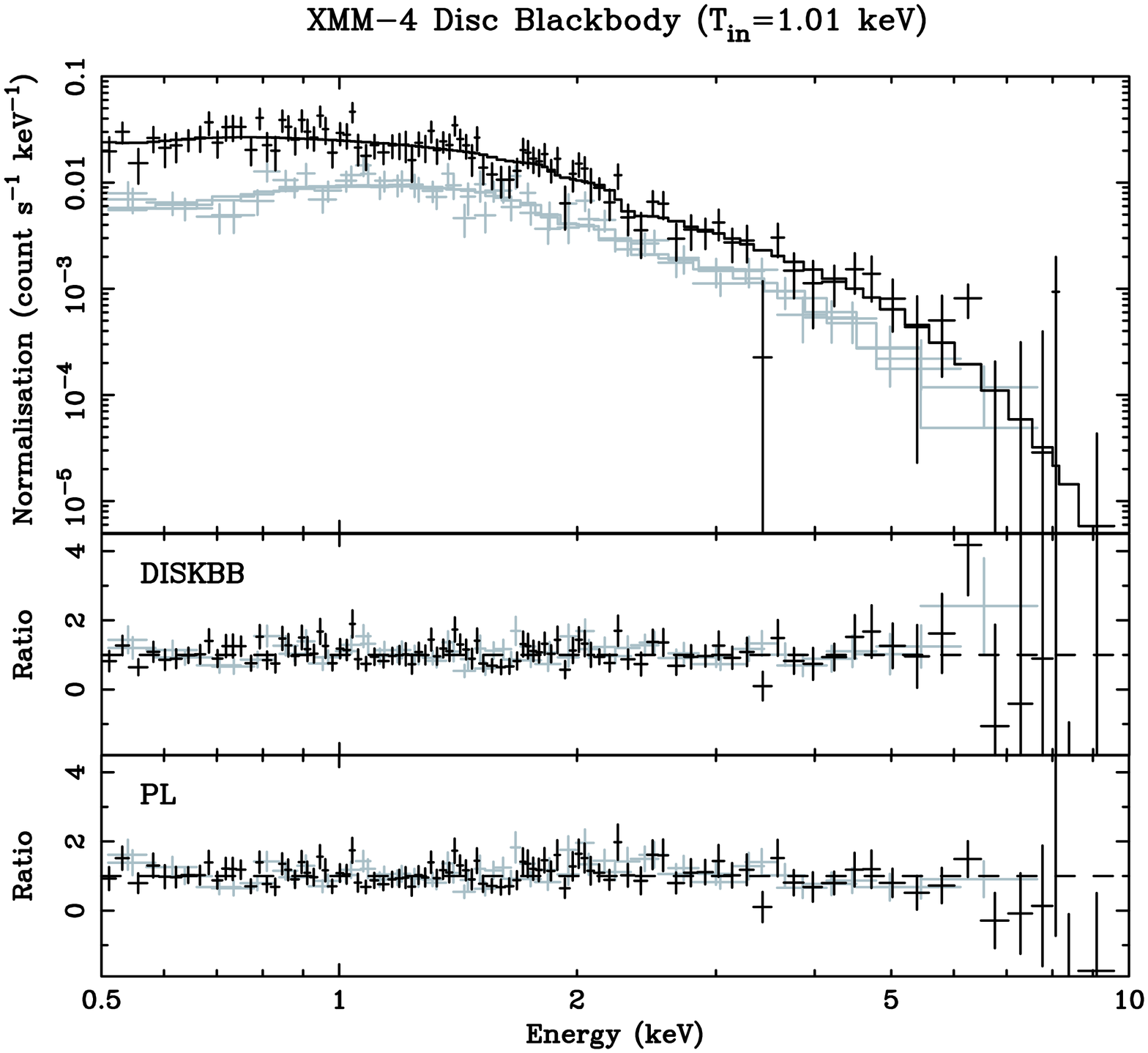}}}
\vspace*{0.5cm}
\parbox{6.5cm}{\scalebox{0.3}{\includegraphics{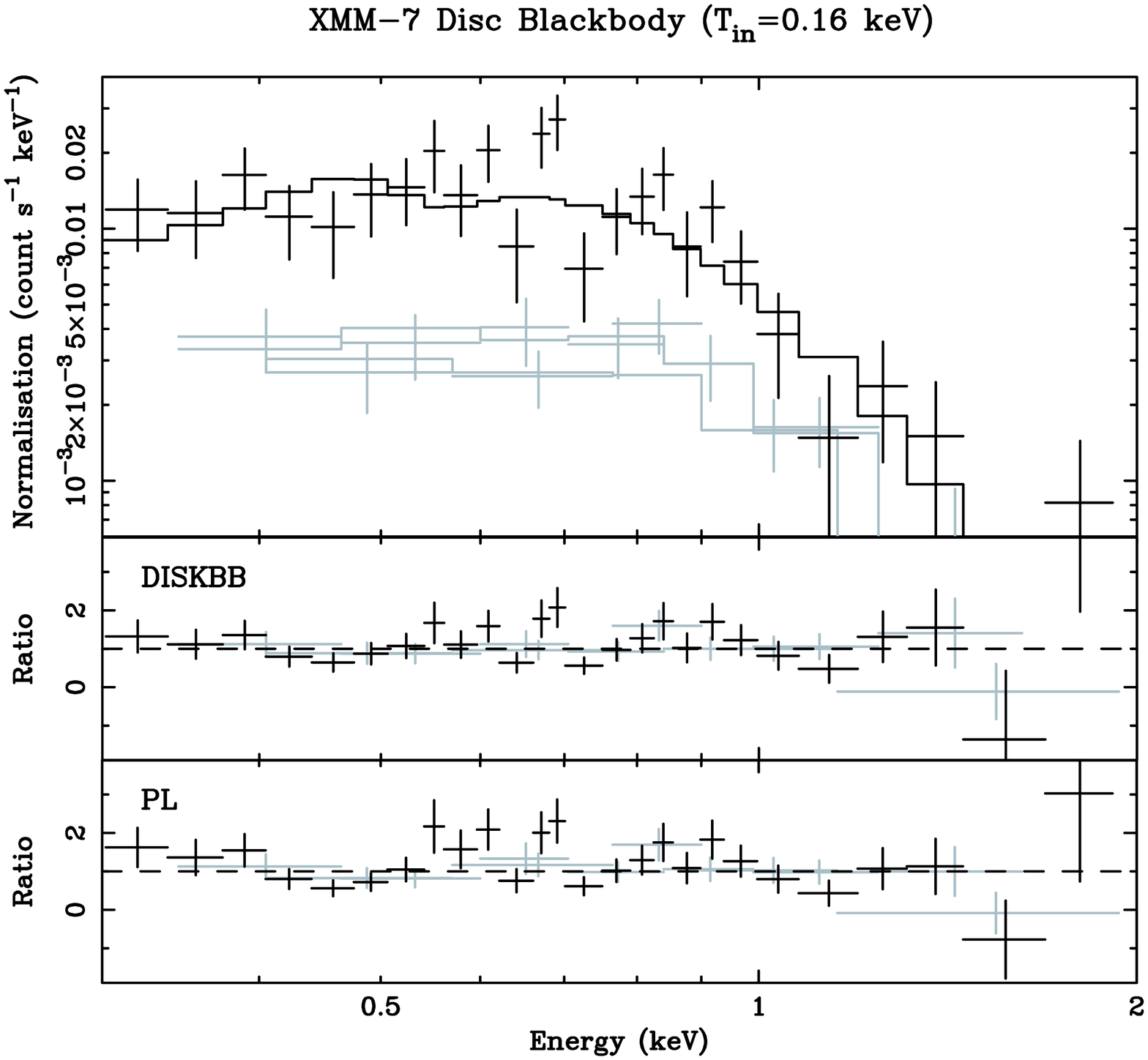}}}
\hspace*{1.5cm}
\parbox{6.5cm}{\scalebox{0.3}{\includegraphics{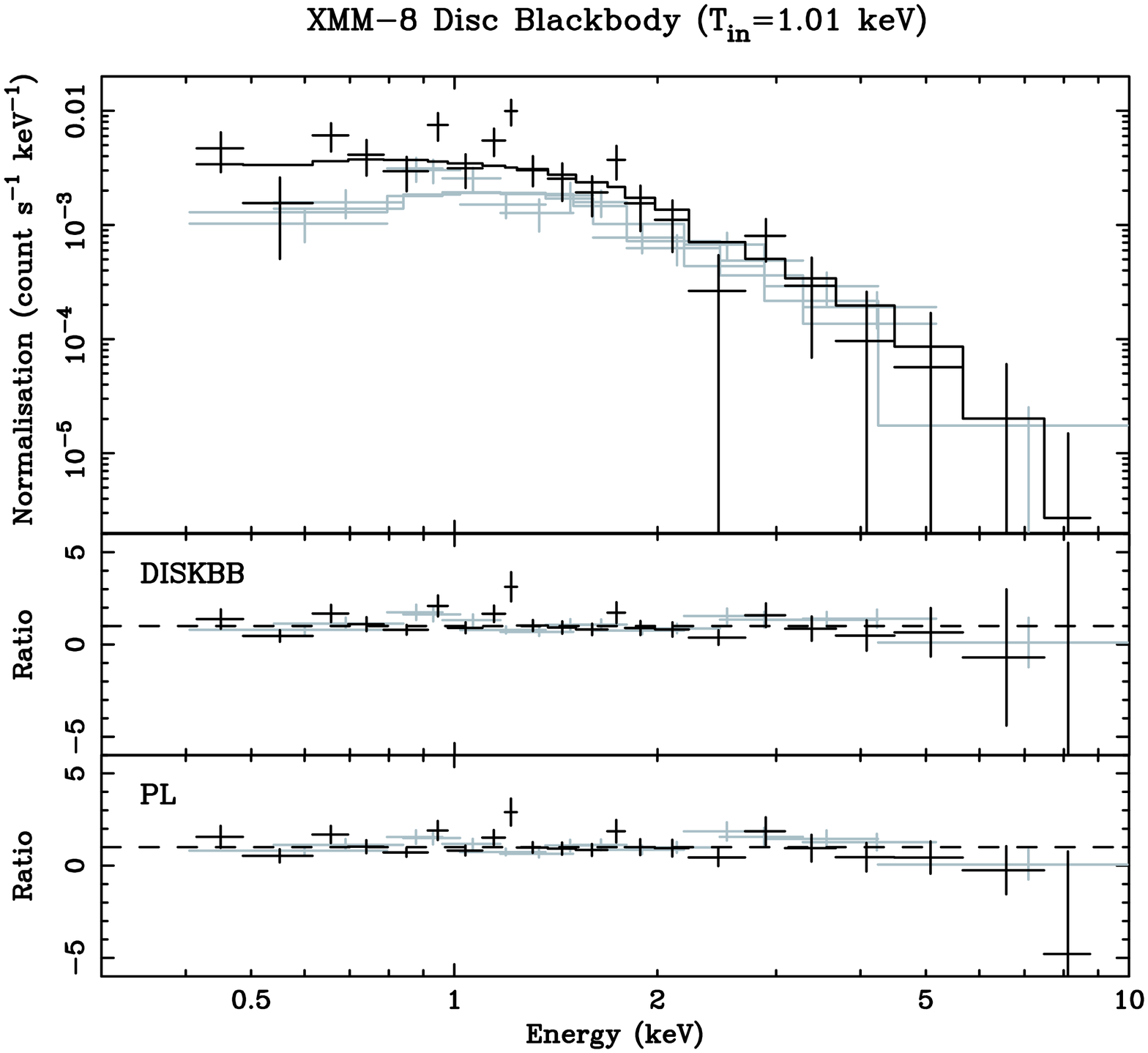}}}
\vspace*{0.5cm}
\parbox{6.5cm}{\scalebox{0.3}{\includegraphics{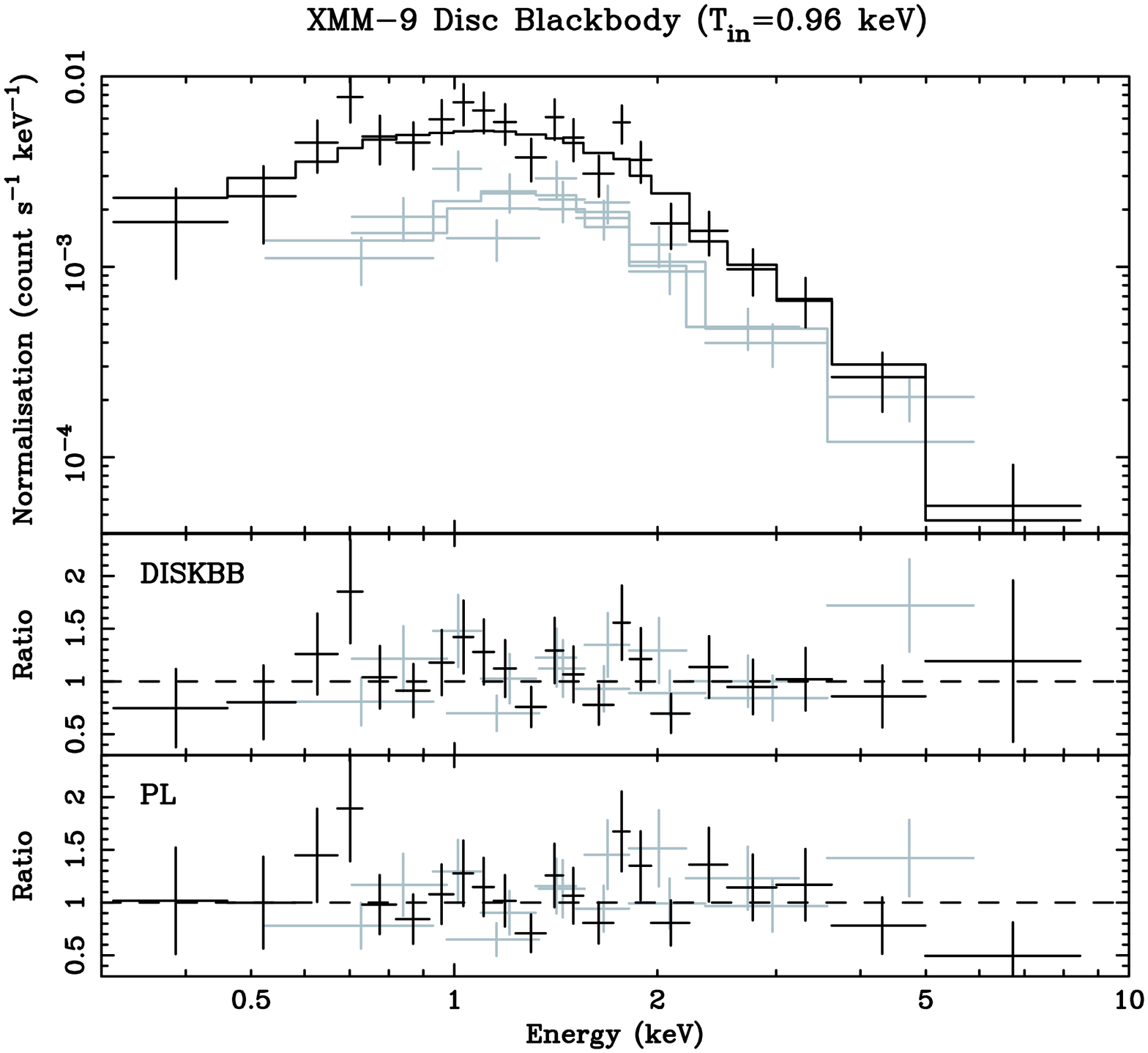}}}
\caption{\xmm spectra and ratios of the disc blackbody sources. PN data points and the best fit model are shown in black; the MOS data are shown in grey.}
\label{fig:spec2}
\end{figure*}

The spectra of eight of the fourteen sources were adequately fit with these simple absorbed one-component models with $\chi^2_\nu\la1$, and their spectra plus best-fit model are plotted in Figures~\ref{fig:spec1} \&~\ref{fig:spec2} together with the ratios from both PL and DISKBB models. In the remaining cases where the $\chi^2_\nu>1$, we went on to fit the spectra with more complex two-component models. The results are shown in Table~\ref{table:spectra2} for the five sources where a significantly better fit at $\ge$ 95 percent confidence level was achieved\footnote{No significant ($\ge$ 95 percent) improvement was made to the fit of XMM-8 with the addition of a soft component.}, with the best-fit model highlighted in bold. Figure~\ref{fig:spec3} shows the spectra/ratios for the best-fit two-component models on the left, while the ratios of the data/model for the remaining fits are shown on the right for comparison. Where the two-component fits did not bring the $\chi^2_\nu$ value down to 1, more complex models were attempted. In only one case did this significantly improve the fit, although the model is partially unconstrained (XMM-13, see section ~\ref{sec:xmm13}). The results of the spectral fitting for each source are discussed in section~\ref{sec:sourceprops}.

For direct comparison of the spectral shapes of the sources between the \chandra and \xmm observations, we have fitted the same set of basic spectral models to the data from the first (long) \chandra observation, which provided sufficient counts for spectral fitting. We have fit the data of ten \chandra sources, nine of which have $>$ 300 counts. The data for CXOU J140314.3+541807 (XMM-2) is below this threshold (178 counts), but we have also fit these data for comparison purposes (although the fit parameters are unconstrained). Standard {\small CIAO} routines were used to extract source and background spectra from the reprocessed event lists and to construct RMF and ARF files. The ARF files were also corrected for the ACIS low energy degradation using the {\it corrarf} tool. The spectral fitting was performed in Sherpa using chi-squared statistics with the Gehrels error approximation, which returns a slightly lower value of the fit statistic as compared with traditional chi-squared statistics.  Errors were calculated for 90 percent confidence using the projection function and Levenberg-Marquardt optimization. The fit results are shown in Table~\ref{table:spectra3}, and all sources are well fit with the single-component models.  However, for comparison purposes, we have also fit the \chandra spectra of the five sources best described by more complex models in the \xmm analysis with the same two-component models. The only statistically significant improvement over a single-component fit is found in the PL+MEKAL fit to CXOU J140414.3+542604 (XMM-3), which we discuss in section~\ref{sec:xmm3}. We compare and discuss the \xmm and \chandra spectral fits in section~\ref{sec:overview}.

\begin{figure*}
\centering
\parbox{6.5cm}{\includegraphics[width=5.2cm,angle=270]{figure5a.ps}}
\hspace*{1.5cm}
\vspace*{0.5cm}
\raisebox{-0.25cm}{\parbox{6.5cm}{\scalebox{0.312}{\includegraphics{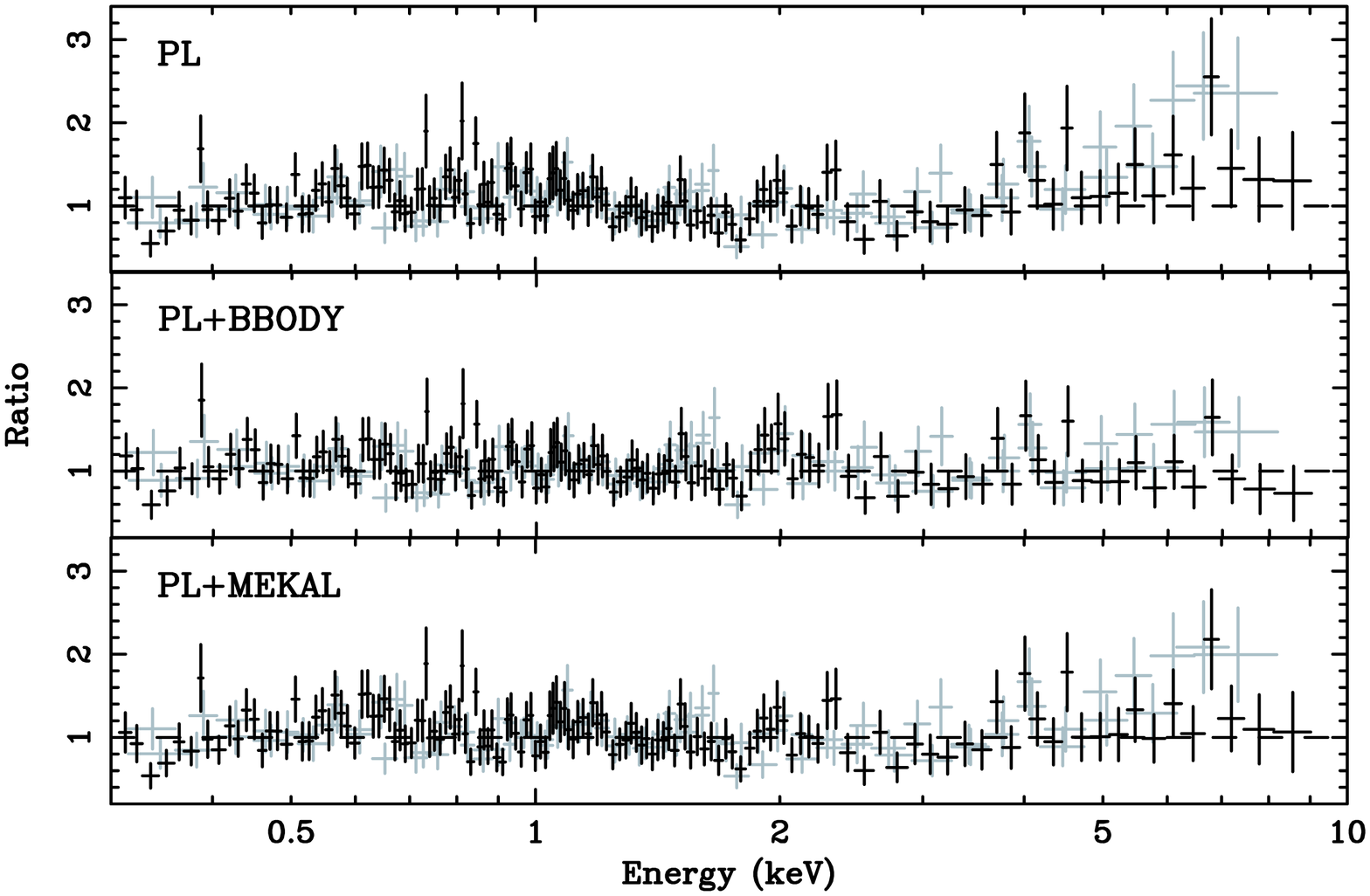}}}}
\vspace*{0.5cm}
\parbox{6.5cm}{\includegraphics[width=5.2cm,angle=270]{figure5c.ps}}
\hspace*{1.5cm}
\raisebox{-0.25cm}{\parbox{6.5cm}{\scalebox{0.312}{\includegraphics{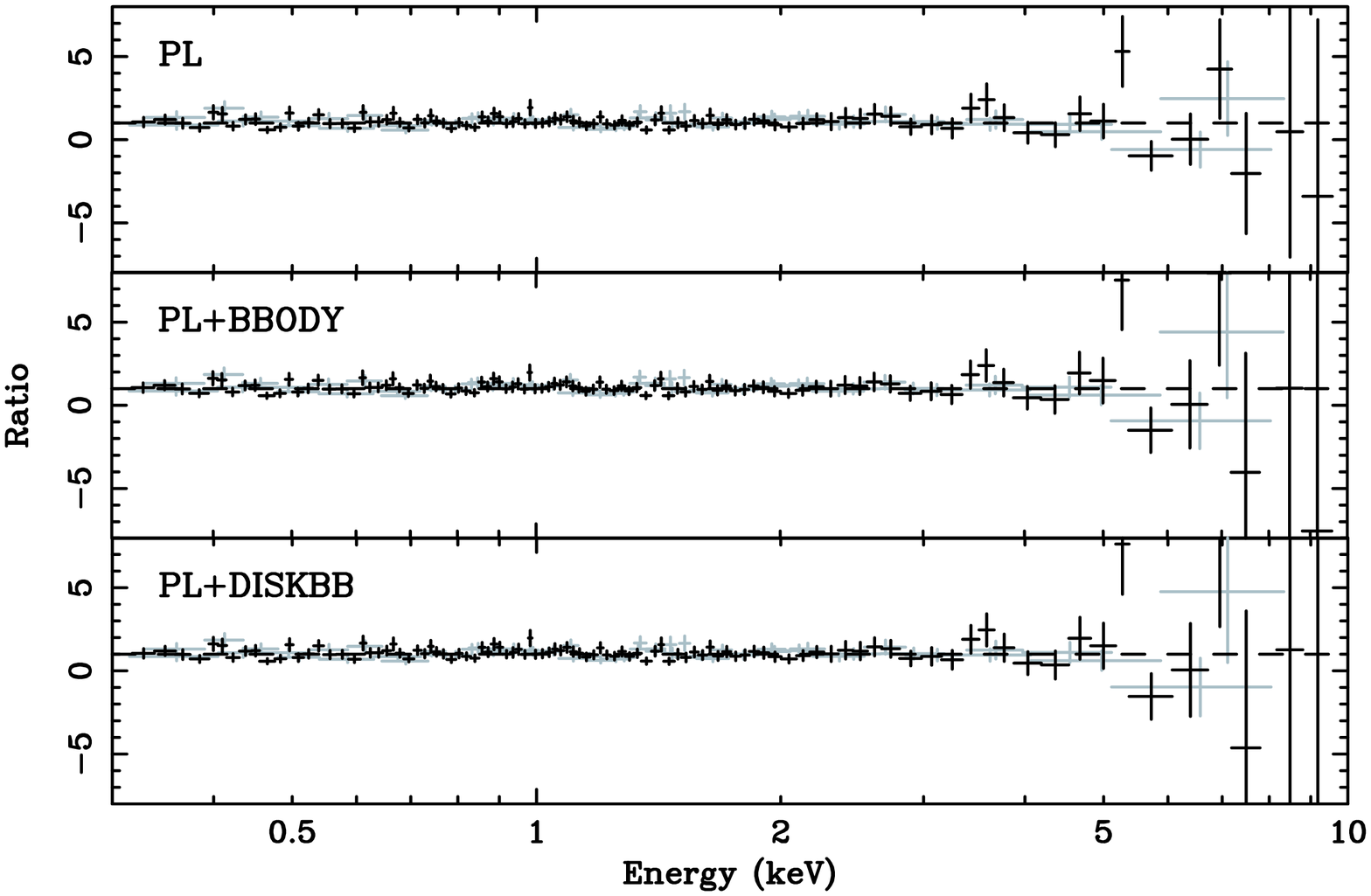}}}}
\vspace*{0.5cm}
\parbox{6.5cm}{\includegraphics[width=5.2cm,angle=270]{figure5e.ps}}
\hspace*{1.5cm}
\raisebox{-0.25cm}{\parbox{6.5cm}{\scalebox{0.312}{\includegraphics{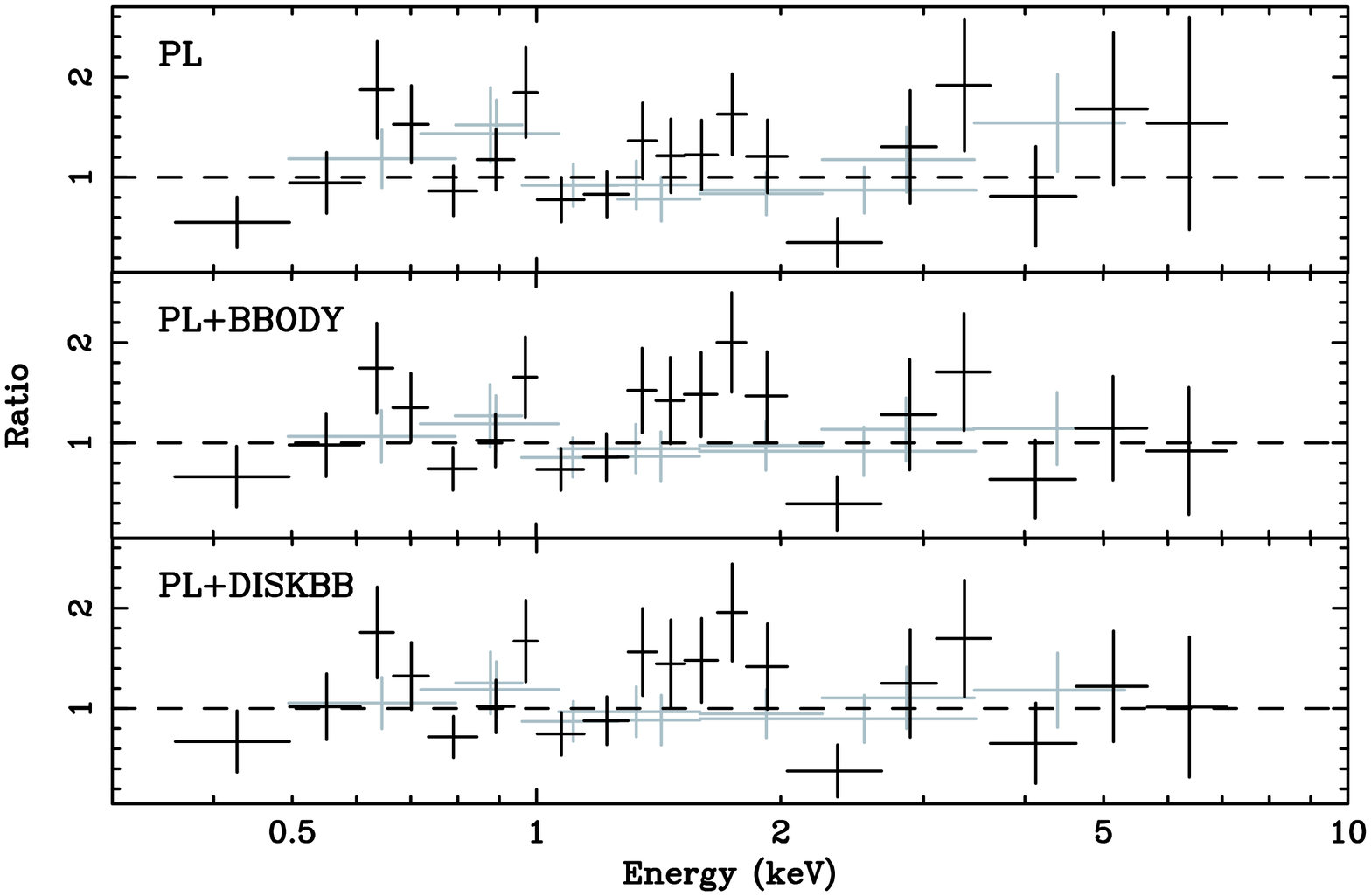}}}}
\vspace*{0.5cm}
\parbox{6.5cm}{\includegraphics[width=5.2cm,angle=270]{figure5g.ps}}
\hspace*{1.5cm}
\raisebox{-0.25cm}{\parbox{6.5cm}{\scalebox{0.312}{\includegraphics{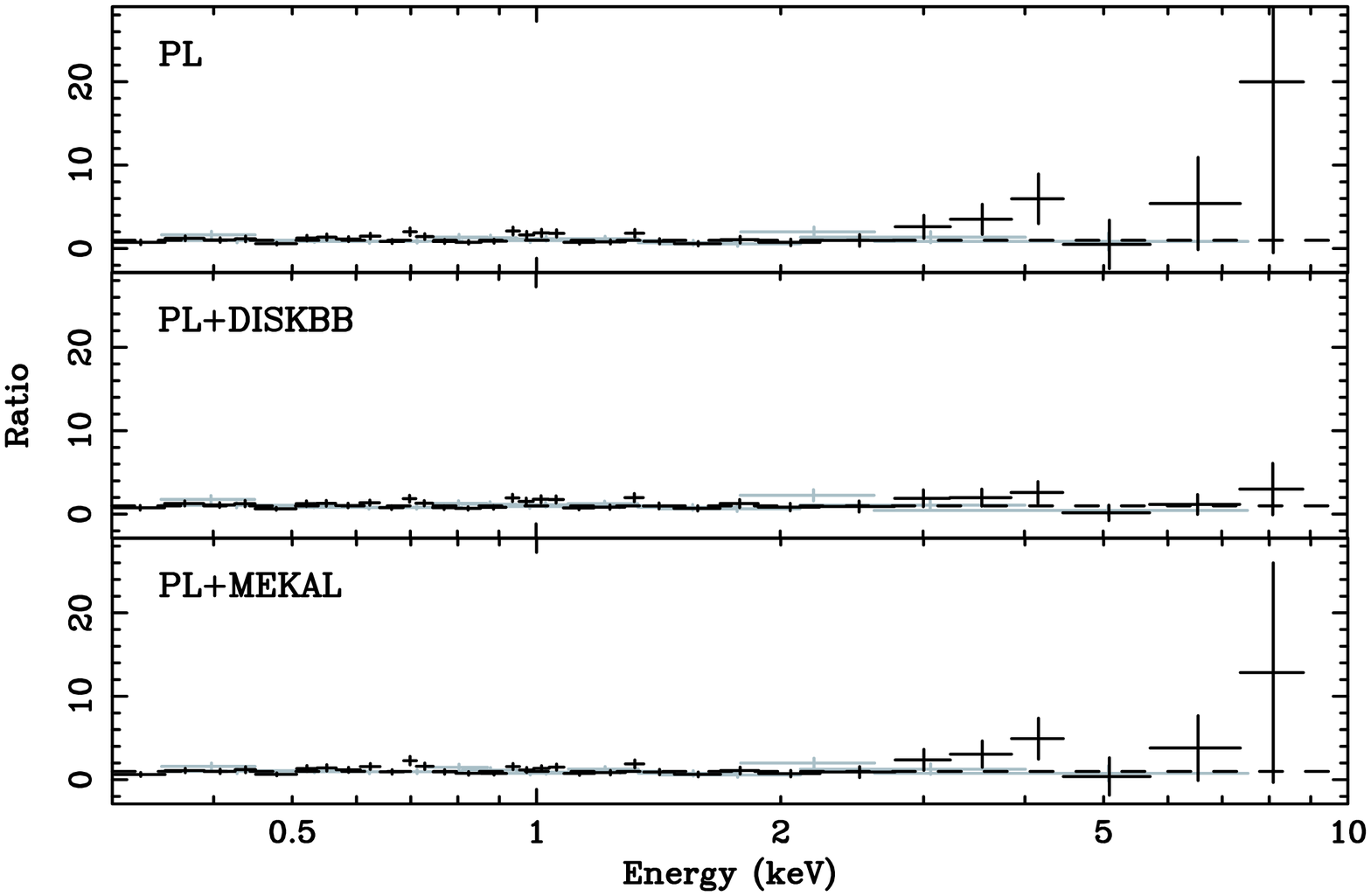}}}}
\caption{[Left] \xmm spectra and ratios of best-fit two-component models. PN data points and the best fit model are shown in black; the MOS data are shown in grey. [Right] Ratios of all other two-component models for the corresponding source on the left.}
\label{fig:spec3}
\end{figure*}

\begin{figure*}
\centering
\hspace*{1.5cm}
\parbox{6.5cm}{\includegraphics[width=5.2cm,angle=270]{figure5i.ps}}
\hspace*{1.5cm}
\vspace*{0.5cm}
\raisebox{-0.2cm}{\parbox{6.5cm}{\scalebox{0.312}{\includegraphics{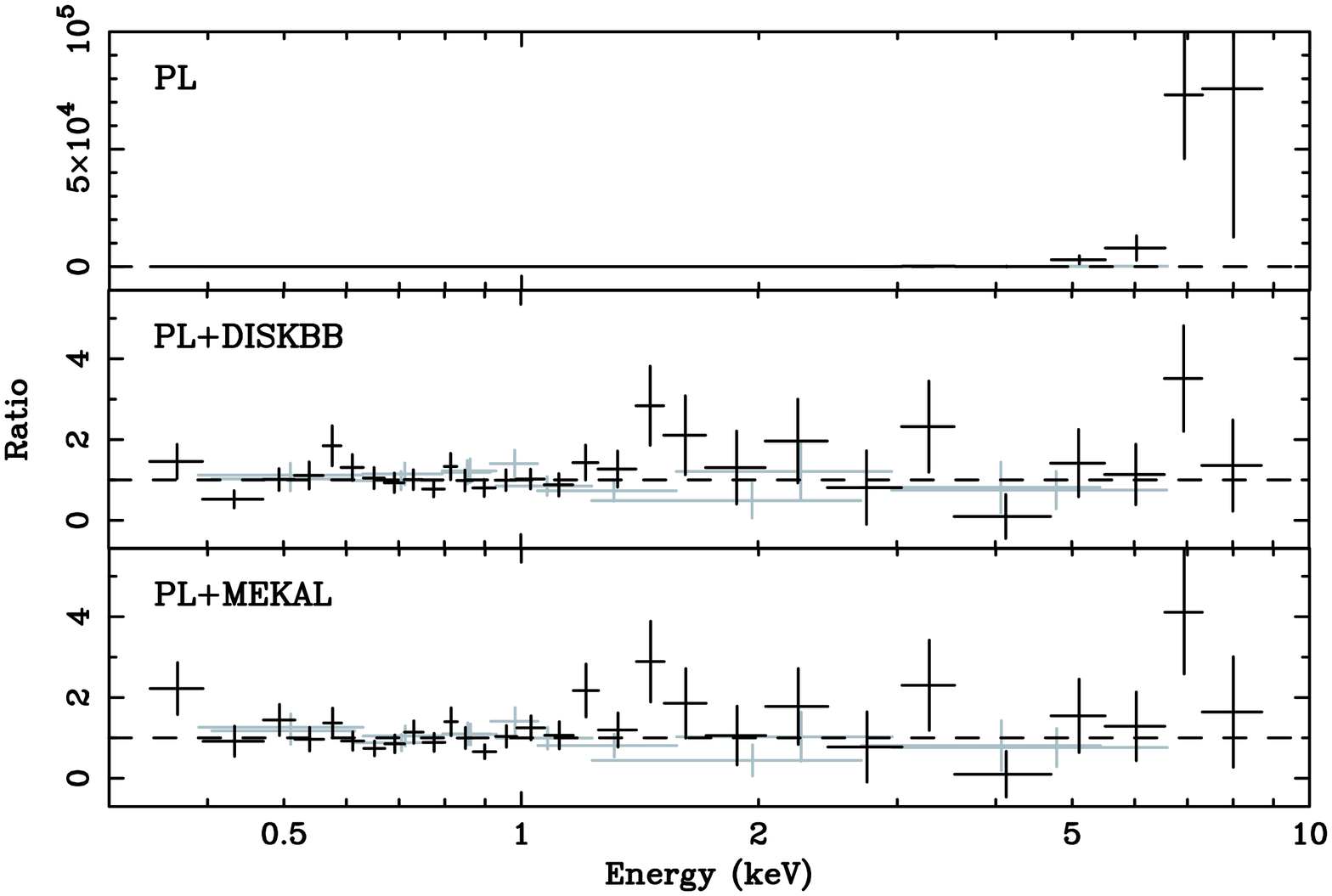}}}}
\contcaption{}
\end{figure*}

\section{Timing Analysis}
\label{sec:timing}

\subsection{Short-term variability}

\begin{figure*}
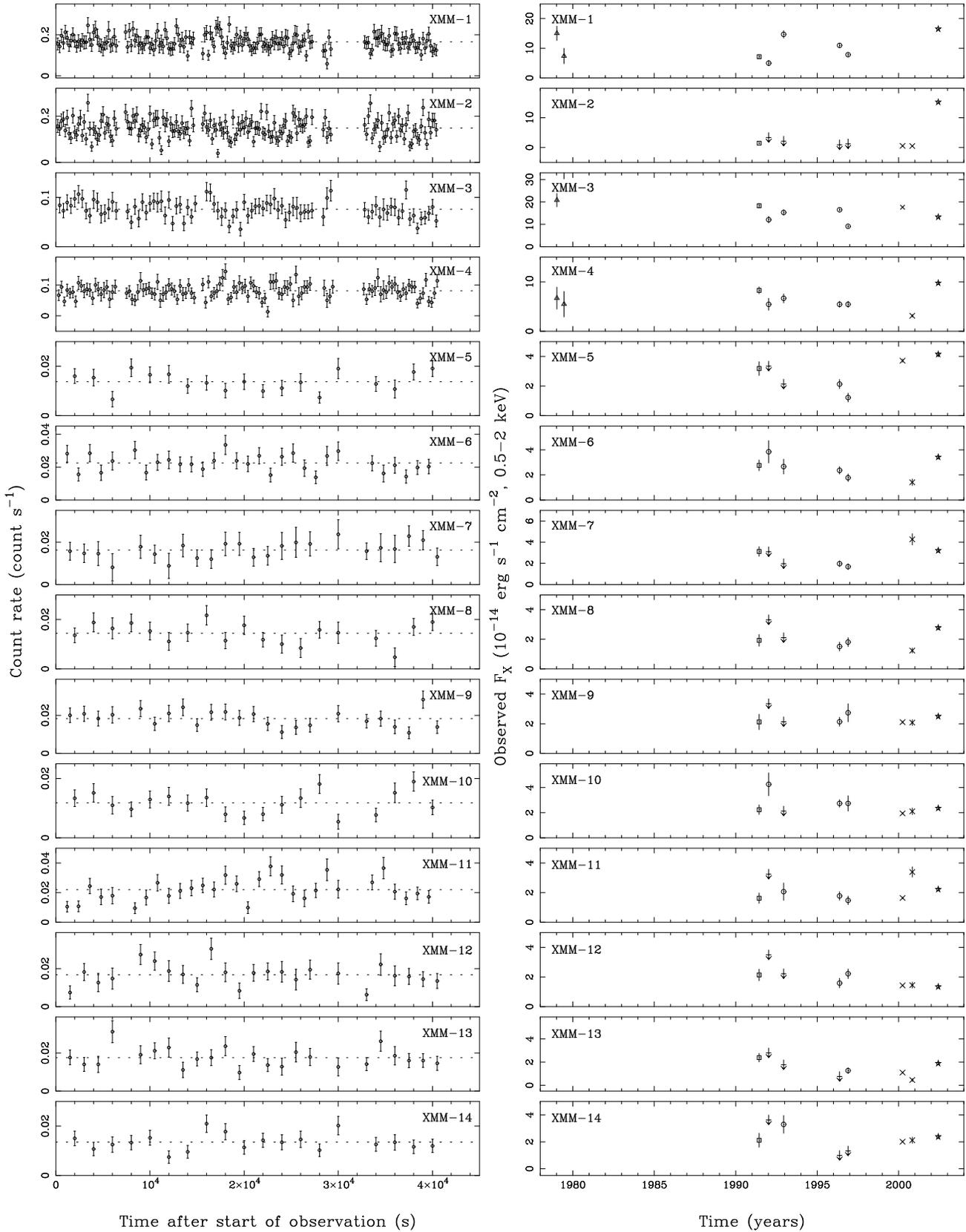

\centering
\scalebox{0.95}{\rotatebox{270}{\includegraphics{figure6a.ps}}}
\scalebox{0.95}{\rotatebox{270}{\includegraphics{figure6b.ps}}}
\caption{[Left] \xmm short-term light curves, obtained by summing the counts from the 3 EPIC cameras. The mean count rate is shown as a dashed line and error bars correspond to 1$\sigma$ deviations assuming Gaussian statistics. [Right] Long-term light curves. The data points shown originate from \einstein IPC (triangles), \rosat PSPC (squares), \rosat HRI (circles), \chandra (crosses) and \xmm (stars) flux measurements, which have been converted to a 0.5--2\,keV flux as described in the text. Arrows denote upper flux limits where no detection was made.}
\label{fig:lcurve}
\end{figure*}

The X-ray variability of each source in the \xmm data was tested by deriving short-term light curves. The data from the three EPIC cameras were co-added to improve the signal-to-noise ratio, and the exposure times were tailored so that each bin had at least 20 counts after background subtraction, giving temporal resolutions ranging from 200-2000 seconds.  The four time intervals of high background flaring were excluded, and we corrected the exposure times and count rates in the incomplete bins adjacent to the gaps with a simple scaling factor, excluding bins with $<$ 0.3 times the exposure remaining of the original bin size. The resulting light curves are shown in Fig.~\ref{fig:lcurve} [left], with error bars corresponding to 1$\sigma$ deviations assuming Gaussian statistics. Two tests were applied to these light curves. Firstly, we calculated a standard deviation of the count rate per bin from the mean count rate and compared this to the expected deviation of $\pm$18--20 percent expected from Gaussian counting noise. Secondly, we performed a $\chi^2$ test to search for large amplitude variability against the hypothesis of a constant count rate. Table~\ref{table:var} shows the results of these tests together with the probability that the data are variable, $P$(var), in the cases where it exceeds 95 percent. These tests on the binned data revealed that ten sources show some evidence of variability, of which four (XMM-2, 4, 11 \& 12) show variability which is significant at the 3$\sigma$ confidence level ($P$(var)$>$99.73). 

In order to search for any indications of gradual small amplitude variations in the source count rates, we conducted Kolmogorov-Smirnov (K-S) tests using PN light curves with a time resolution of 1 second\footnote{Only the PN data were used for the K-S test as the MOS cameras are limited to 2.6 second time resolution in Full Window mode.}. For each source, we compared the observed background-subtracted cumulative photon arrival distribution with the expected distribution if the flux was constant. Of the fourteen sources, four show evidence for variability, but only one (XMM-11) is statistically variable at the 3$\sigma$ confidence level. Initially, source XMM-7 also exhibited variability in the K-S test, but further investigation showed that this could be attributed to variability in the underlying background flux in this region of the detector, which constituted 57 percent of the total (source+background) flux for this source.

Similar tests on the \chandra data detect significant short-term variability only in CXOU J140336.0+541925 (XMM-11), with $P$(var)$>$99.9 in both the $\chi^2$ and K-S tests.

\subsection{Long-term variability}

We have taken advantage of the fact that M101 is a well-studied galaxy that has been observed in the X-ray regime with the {\it Einstein}, {\it ROSAT}, \asca and \chandra observatories. We have used these data (except \asca due to its poor angular resolution) to construct long-term light curves for the fourteen sources.  For consistency between the datasets, we have plotted the fluxes in terms of the observed 0.5--2\,keV flux, as this is the energy range covered by all the missions. The \xmm and \chandra fluxes are measured directly from the best-fit absorbed single component models. We have used {\small webPIMMS} to derive fluxes for the \einstein IPC and \rosat PSPC/HRI observations using the observed count rates, assuming the absorbed PL continuum slope measured in the \xmm observation and normalized to the 0.5--2\,keV band. In the two \einstein IPC observations \citep{trinchieri90}, the central region of the galaxy is not resolved, but emission is detected from three of the \xmm sources away from the centre of the galaxy (XMM-1, 3 \& 4). All fourteen sources were detected in the \rosat PSPC observation, and our fluxes are derived from the count rates given in \citet{wangetal99}. However, since the \rosat HRI count rates quoted in \citet{wangetal99} are a combination of four individual observations, we have re-analysed the archival data (as per \citealt{roberts00}) and measured count rates (or upper limits) for each individual data set. Finally, we correct the count rate of the \rosat PSPC source P19, which is resolved into two sources by both the \rosat HRI (H29 \& H30) and \xmm (XMM-9 \& 14). Since the fluxes of these two sources are almost identical in the first HRI observation, we have attributed half of the PSPC flux to each. Note also that as XMM-2 and XMM-10 are resolved into two sources in the \chandra observations, the \chandra data points represent their combined flux. 

The resulting light curves are shown in Figure~\ref{fig:lcurve} [right], and although poorly sampled with only 8--10 observations in 11--24 years, the majority of the sources do show evidence of long-term variability/transient behaviour. These results are discussed in section~\ref{sec:overview}.

\section{Individual Source Properties}
\label{sec:sourceprops}

In this section, we discuss the spectral and timing properties of each source in turn. For consistency, we quote the luminosities for both the \xmm and \chandra data in the 0.3--8\,keV band. An overview of the spectral and temporal properties for all sources is given in section~\ref{sec:overview}.

\subsection{XMM-1}
\label{sec:xmm1}

This source is spatially coincident with the SNR MF37 in an outer spiral arm of the galaxy. However, the \rosat data showed evidence of strong variability, implying that it is a single accreting object, perhaps an XRB formed in the supernova explosion. It is the brightest source in the \xmm observation, and its spectral shape (disc black-body model with $T_{in}$=1.33\,keV) plus indication of short-term variability strongly support the hypothesis that this is a single XRB. Its long-term lightcurve also shows significant variability of a factor of $\sim$3--4 between observations, with an increase in the \xmm observation putting it well into the ULX regime (intrinsic $L_X=3.3\times10^{39} \ergsec$). Similar high inner accretion disc temperatures have been observed in other ULXs, e.g. the \asca study of \citet{makishima00}. This is the only source that was not covered by either of the \chandra observations.

\subsection{XMM-2}
\label{sec:xmm2}

This source is located in a spiral arm region containing numerous \hii regions. It is coincident with the \rosat source H25(P17), but is resolved into two sources (CXOU J140313.9+541811 \& CXOU J140314.3+541807) in the \chandra observations. However, only CXOU J140314.3+541807 has sufficient counts for spectral fitting and it is marginally best fit with an (unconstrained) absorbed PL model ($\Gamma$=2.4) with an unabsorbed X-ray luminosity of $L_X=7.9\times10^{37} \ergsec$, although a disc blackbody model gives a statistically indistinguishable fit ($T_{in}$=0.6\,keV).  The second source (CXOU J140313.9+541811) has a luminosity of $L_X=3.4\times10^{37} \ergsec$ if we assume the spectral fit parameters of the \xmm PL model described below.

Although we expect that the emission we detect in the \xmm observation is the unresolved flux from both \chandra sources, XMM-2 is $\sim$30 times more luminous than either of the \rosat or two \chandra sources, and the peak of the emission corresponds more closely to the position of the brighest \chandra source (CXOU J140314.3+541807, offset $\sim$1.8 arcseconds). The \xmm spectrum of this source is best described by a two-component model (see Tables~\ref{table:spectra1} \&~\ref{table:spectra2}). The spectral fit is improved over that of a simple PL fit with the addition of either a DISKBB, BBODY or MEKAL component. The PL+DISKBB is statistically the best fit, but it is almost indistinguishable from the PL+BBODY fit with similar temperatures and PL slopes.   The short-term light curve of this observation shows definite variability, indicating that it is dominated by a single accreting source. Its unabsorbed luminosity in this observation is $L_X=2.7\times10^{39} \ergsec$, classing it as a ULX. The sharp increase in luminosity in the \xmm data compared with the previous observations leads us to conclude that this is an X-ray transient source which is likely to be related to the active star formation in its vicinity.

\subsection{XMM-3}
\label{sec:xmm3}

Unlike the previous sources, this is not associated with a region of obvious star formation, but it is located in a region in the direction of the end of a spiral arm (Fig.~\ref{fig:dssim}). The \rosat data showed that it was strongly variable by a factor of $\sim$2 on timescales of days. The \chandra data are well fit with a soft PL continuum ($\Gamma$=3.6), which yields a high unabsorbed luminosity ($L_X=1.7\times10^{39}\ergsec$ observed; $L_X=1.2\times10^{40} \ergsec$ instrinsic), although there is a large uncertainty associated with the intrinsic luminosity due to the large absorption correction ($N_H=3.5\times10^{21}$ cm$^{-2}$).  It is also in the ULX regime in the \xmm observation ($L_X=3.1\times10^{39} \ergsec$), and the spectrum is best described with an absorbed PL+MEKAL thermal plasma model ($\Gamma$=2.6, $kT$=0.98\,keV). Even though the $\chi^2_\nu$=1.05 for this spectral model, we tested whether the fit would be improved by adding an extra MEKAL component, but this made the fit marginally worse. However, if we fit the \chandra spectrum with a PL+MK model we get a significantly improved fit ($N_H$=3.23$^{+0.38}_{-0.36}$, $kT$=0.80$^{+0.22}_{-0.16}$\,keV, $\Gamma$=3.42$^{+0.19}_{-0.18}$, $\chi^2_{\nu}\sim0.9$) which gives additional evidence of the presence of thermal plasma in the vicinity of this source. We detect no definite variability in its short-term light curve, but its long-term light curve shows that it has varied by a factor of $\sim$4 between observations spanning $\sim$24 years. This variable behaviour shows that this source is almost certainly dominated by an XRB. Although the MEKAL component only contributes 5 percent of the total flux of this source, it has a high luminosity of $L_X\sim1.6\times10^{38} \ergsec$. The source is too far away from the centre of the galaxy to be contaminated by diffuse emission from the disc. Neither is it in a star-forming region of M101 and no SNR has been detected at this position. We therefore speculate that the thermal plasma could originate in a hot photoionized plasma (e.g. stellar wind) surrounding the XRB, if the companion is a high-mass star (see section~\ref{sec:2comp}).

\subsection{XMM-4}
\label{sec:xmm4}

This source is located in the giant \hii complex NGC~5447 in a spiral arm of M101. \citet{wangetal99} originally classed it an XRB, based on the hard spectral shape implied by the \rosat PSPC harness ratios. This source was detected in the short second \chandra observation, although it has too few counts for spectral fitting.  This source has brightened considerably (factor of $\sim$3) in the \xmm observation ($L_X=1.3\times10^{39} \ergsec$), and its spectra are best fit with absorbed DISKBB model with a high disc temperature ($T_{in}=1.0$\,keV), again similar to the ULXs measured by \citet{makishima00}. Both the short-term \xmm and long-term light curves show evidence of variability; this coupled with its spectral shape leads us to conclude that the emission is due to an XRB embedded in the \hii region.

\subsection{XMM-5}
\label{sec:xmm5}

This source is located away from the main body of the galaxy and was classed as an AGN by \citet{wangetal99} according to its hardness ratios, the fact that it has a blue stellar counterpart and lack of short-term variability. It is however in the direction of an outer spiral arm and may be a source associated with M101. In the \chandra observation, its spectrum is well fit with an absorbed PL ($\Gamma$=1.8). The best fit single component model to the \xmm data is an absorbed PL with an unabsorbed luminosity of $L_X=9.1\times10^{38} \ergsec$, but the introduction a cool MEKAL component does improve the fit ($\Gamma$=2.1, $kT$=0.17\,keV). The MEKAL component contributes a substantial amount (20 percent) to the total flux from this source, which could indicate the presence of substantial star formation activity in this region, or might again be attributable to the source itself. Its long-term light curve shows a factor of $\sim$3 variability between the different observations, and there is some evidence of short-term variability in the \xmm data. This raises the possibly that it is an accreting XRB in M101, although we cannot discount the possibility that this is a background AGN, as rapid X-ray variability on the timescales of hours is known to exist some in BL Lac objects (e.g. \citealt{tagliaferri03}).  According to the USNO-A2 catalogue, the apparent $B$ magnitude of the optical counterpart is 18.7, which, assuming a $F_{\nu}\sim\nu^{-1}$ spectrum typical for an AGN, equates to an apparent $R$ magnitude of $\sim$17.8. Using the 2--10\,keV flux of $6.6\times10^{-14} \ergcms$, this in turn equates to an X-ray-to-optical flux ratio $log(F_X/F_R)$=-0.56, which is within the range measured for optically identified AGN in the \xmmn hard X-ray survey of \citet{piconcelli03}.  However, such objects do not typically show MEKAL components in their X-ray spectra. Alternatively, if the optical source is located in the outer spiral arm of M101, the implied absolute magnitude of $M_B=-10.6$ is consistent with the range measured in the $V$-band for stellar clusters in the Antennae ($-9>M_V>-14$, \citealt{whitmore99}). Therefore, although this source may be a background AGN, its short- and long-term variability plus the presence of thermal plasma in its vicinity makes it much more likely that this is an XRB associated with star formation activity in M101.

\subsection{XMM-6}
\label{sec:xmm6}

This source is located in the same spiral arm as XMM-4 close to an \hii region.  Although no short-term variability is detected in the \rosat HRI data, this source showed hard X-ray spectral characteristics in the PSPC data which lead the authors to conclude that it is an XRB. It was detected in the short \chandra observation, with an unabsorbed luminosity of $L_X=3.3\times10^{38} \ergsec$. The \xmm spectrum is best fit with a soft absorbed PL model ($\Gamma$=2.6) with an intrinsic luminosity of $L_X=7.9\times10^{38} \ergsec$. Its short-term light curve doesn't show any indication of variability, although its long-term light curve does. The observed flux peaks in the first \rosat HRI observation, gradually fades up to the \chandra observation and sharply increases in the \xmm observation by a factor of $\sim$3. This variable behaviour leads us to conclude that this source is an XRB.

\subsection{XMM-7}
\label{sec:xmm7}

This source is one considered by \citet{wang99} to be a HNR candidate. The \rosat HRI source (H49) is positionally coincident within 1$\sigma$ error radius of the optical position of NGC~5471B, one of the SNRs detected in NGC~5471 by \citet{skillman85} in the radio and \citet{chu86} in the optical. \citet{wang99} found that the \rosat PSPC data was consistent with a Raymond-Smith thermal plasma with a temperature of 0.29\,keV and an X-ray luminosity of $L_X=3\times10^{38} \ergsec$ (0.5--2\,keV). Using these characteristics, they employed the Sedov solution to estimate the total thermal energy of the SNR to be $\sim10^{52}$ ergs (if that is where the X-ray emission originates). This is ten times that expected from a typical supernova and so have suggested that it could be a HNR. No temporal variability was detected in the \rosat data, consistent with the hypernova scenario.

More recently, \citet{chen02} have studied NGC~5471B using \hst (HST) WFPC2 contiunuum band and emission lines image together with echelle spectra for kinematic studies. They found three [SII]-enhanced shells in NGC~5471, the brightest of which is NGC~5471B with an expansion velocity of $\geq$ 330 km s$^{-1}$, a kinetic energy of $5\times10^{51}$ ergs and hence an explosion energy $>10^{52}$ ergs.  The WFPC2 images showed that it has a complex environment of many concentrations of stars and that the shell itself encompasses two young OB associations. The \rosat HRI contours are centered on NGC~5471B on the HST images and the authors conclude that the SNR within it was produced by a hypernova since a shell moving at that velocity is unmatched by any comparable shells in other galaxies that have been produced by multiple supernovae.

This source is an off-axis detection in the short \chandra observation, and hence has insufficient counts for spectral fitting or an accurate determination of its position. The \xmm spectrum shows that this source is very soft. It is well fit with both an absorbed cool BBODY ($T_{in}$=0.14\,keV, $\chi^2_\nu$=0.93) model with an unabsorbed luminosity of $L_X=5.2\times10^{38} \ergsec$ and a DISKBB ($T_{in}$=0.16\,keV, $\chi^2_\nu$=0.96) model with an unabsorbed luminosity of $L_X=1.0\times10^{39} \ergsec$. This cool temperature classifies this source as a luminous supersoft source (SSS), similar to those found in other galaxies (e.g. \citealt{distefano03a}). Although not shown in Table~\ref{table:spectra1}, this source was also adequately fit with an absorbed single MEKAL plasma model ($kT=$0.1\,keV, $\chi^2_\nu$=1.2). Its light curve shows no variability over the timescale of the \xmm observation, although there is evidence of variability in the long-term light curve (factor of $\sim$2.5). The nature of this source therefore remains open to debate. A reasonable MEKAL X-ray spectrum and a lack of variability are still consistent with an interpretation as a HNR. However, the similarity of its spectral form to other SSSs, suspected to be XRBs, offers an alternative solution.

\subsection{XMM-8}
\label{sec:xmm8}

XMM-8 is located away from the main body of the galaxy, but it is in the direction of one of the outer spiral arms (see Fig.~\ref{fig:dssim}). It was detected in the \rosat observations, but showed no variability. It was also detected in the short \chandra observation, although with insufficient counts for spectral fitting.  However, the \xmm spectra are best fit with an absorbed DISKBB ($T_{in}=1.01$\,keV) with an unabsorbed luminosity of $L_X=3.9\times10^{38} \ergsec$. Even though this fit has a $\chi^2_\nu$ value of 1.33, the addition of a second component does not significantly improve the fit. There is some evidence of variability in the short-term \xmm light curve, and the long-term data show a factor of $\sim$2 variability between observations, with a sharp increase in flux in the \xmm observation. We therefore conclude that this source is likely to be an XRB.

\subsection{XMM-9}
\label{sec:xmm9}

This source was orginally identified by \cite{wang99} as being possibly coincident with SNR MF54. Its \rosat X-ray position was offset from the optical position of MF54 by 6.2 arcseconds (2$\sigma$ error radius) and therefore considered to be a possible HNR due to its high X-ray luminosity ($L_X=1.4\times10^{38} \ergsec$) and lack of temporal variability. However, this source was detected in the long \chandra observation, and it was clearly demonstrated by \citet{snowden01} that its position (accurate to $\sim$0.5 arcseconds) is inconsistent with the optical position of MF54 with an offset of 8.3 arcseconds between them. They did detect a faint X-ray source (P67) at the position of MF54 with an X-ray luminosity of $L_X=4.3\times10^{36}$ ergs s$^{-1}$ (thermal model with $kT=0.73$) which is reasonable for a SNR with a diameter of $\sim$20 pc as determined by \citet{matonick97}. Our analysis of the \chandra data for the brighter source show that its spectrum is best fit with a hot DISKBB model ($T_{in}$=1.1\,keV) with an unabsorbed X-ray luminosity of $L_X=3.4\times10^{38} \ergsec$, although it is also well fit with an absorbed PL ($\Gamma$=2.0).

The \xmm spectra are consistent with this scenario, being best fit with an DISKBB model with a similar temperature ($T_{in}$=0.96\,keV) with an unabsorbed luminosity of $L_X=4.2\times10^{38} \ergsec$. Its short-term \xmm light curve shows no indication of variability, and likewise its long-term lightcurve appears fairly constant over the 11 year period. However, its spectral shape leads us to conclude that it is an accreting XRB, with an inner disc temperature similar to those ULXs observed with {\it ASCA}.

\subsection{XMM-10}
\label{sec:xmm10}

This is the source located at the optical nucleus of M101. In the \rosat data it has an extended morphology and shows no variability. In the \chandra observation, the nucleus is resolved into two point sources (CXOU J140312.5+542053 \& CXOU J140312.5+542057), both of which are best fit with absorbed PL models ($\Gamma$=1.7/2.1). \citet{pence01} have shown that the most northerly source (CXOU J140312.5+542057) coincides with the nucleus, the other (CXOU J140312.5+542053) coincides with a cluster of bright stars 3.1\arcs ($\sim$110 pc) to the south. Neither source shows temporal variability and they have similar unabsorbed luminosties of $L_X=1.3\times10^{38} \ergsec$ (CXOU J140312.5+542053) \& $L_X=9.3\times10^{37} \ergsec$ (CXOU J140312.5+542057).  

The two sources are not resolved in the \xmm data, but the properties of XMM-10 are fairly consistent with their combined emission. The spectra are best fit with an absorbed PL model ($\Gamma$=2.3) with a luminosity of $L_X=3.3\times10^{38}$ ergs s$^{-1}$. However, the \xmm short-term light curve does show some evidence for variability over the duration of the observation. The long-term lightcurve shows that the combined flux from these two sources has remained relatively constant over the 1996-2002 period, but there is some variability apparent in the early data (factor of $\sim$2), indicating that at least one of the two sources is a variable XRB system.

\subsection{XMM-11}
\label{sec:xmm11}

This is another of the five X-ray sources considered by \citet{wang99} to be a HNR candidate based on its positional coincidence with SNR MF83 and its high X-ray luminosity ($L_X=1.2\times10^{38}$ ergs s$^{-1}$). With the \rosat HRI and PSPC data, \citet{wang99} was unable to constrain the shape of its X-ray spectrum. However, he comments that if the X-ray emission is diffuse (i.e. a thermal plasma), then its X-ray luminosity suggests a total thermal energy of $\sim6\times10^{52}$ ergs. The HNR scenario was supported by the lack of temporal variability in their data.

MF83 is located between two spiral arms (see Fig.~\ref{fig:dssim}), and is one of the largest SNRs identified by \citet{matonick97} in M101 by its high [SII]/H$\alpha$ ratio and bright [OIII] emission, clearly suggesting that the optical nebula is shock heated. However, based on ground-based and HST optical observations, \citet{lai01} showed that MF83 is a star-forming region consisting of a $\sim$270\,pc ionized gas shell with an expansion velocity of $\sim$50 km s$^{-1}$. It has composite sources at its centre which are likely to be OB associations and four \hii regions along its rim. The large size of the shell and the fact that it is centered on groups of stars suggest that this shell is a superbubble similar to those found in the Large Magellanic Cloud, rather than a SNR.

The \chandra data have been able to constrain the X-ray spectrum of this source. It is best fitted by an absorbed PL ($\Gamma$=2.7), and is poorly fitted by thermal emission model. Significant short-term variability was found in the \chandra light curve ($\chi_{\nu}^2\sim$4.3), which rules out the hypernova scenario. The variability, together with the PL spectral shape suggests that the origin of the X-ray emission is likely to be an XRB associated with the superbubble. The position of the \chandra source is coincident with the MF source position to $\sim3$ arcseconds.

The \xmm spectra of this source are not adequately fit with a simple one-component model, but the only two-component model which shows a significant improvement is the PL+BB ($\Gamma$=1.7, $kT=0.19$\,keV).  The PL+DISKBB fit ($\Gamma\sim$1.7, $T_{in}=0.22$\,keV) gives an improvement at the 93 percent significance level, with parameters similar to IMBH candidates studied by \citet{miller03a}. Its short-term \xmm light curve also shows significant variability, supporting the idea that this is not a HNR. The long-term lightcurve also shows some variability (factor of $\sim$2). It has an unabsorbed luminosity of $L_X=6.2\times10^{38} \ergsec$ and we conclude that this source is an XRB associated with the superbubble.

\subsection{XMM-12}
\label{sec:xmm12}

This source wass classed as an `interarm' AGN by \citet{wangetal99} based on its hardness ratios, blue optical counterpart and lack of temporal variability in the \rosat HRI data. The \chandra data are well fit with either an absorbed PL model ($\Gamma$=1.5) or a DISKBB model ($T_{in}$=1.61\,keV) and no variability is evident. However, \citet{mukai03} do note that the \chandra position is 5 arcseconds from the optical position of the blue counterpart reported by \citet{wangetal99} and so the source may indeed be located within a spiral arm in M101. 

The \xmm data are also best fit with an absorbed PL model ($\Gamma$=1.8) with an unabsorbed luminosity  of $L_X=4.1\times10^{38} \ergsec$, although the DISKBB also gives a very good fit to the data ($T_{in}=1.45$\,keV).  Its long-term light curve appears fairly constant, but its short-term lightcurve shows strong evidence for variability, leading us to conclude that it is an XRB.
\\

\subsection{XMM-13}
\label{sec:xmm13}

This source is coincident with NGC~5461, another of the bright giant \hii regions in M101. Thermal and non-thermal radio emission has been detected in NGC~5461 \citep*{graeve90}, and the kinematic study by \citet{chu86} revealed the presence of high-velocity (455\,km s$^{-1}$ FWZI) gas near its centre. However, no SNR was detected at that position by \citet{matonick97}. The presence of high-velocity gas could be explained by a wind-blown bubble around stars embedded in the \hii region, although the velocities measured here are far greater than those detected in other galaxies such as the LMC \citep{chu86}.

The \rosat observations characterized this source as diffuse, with no temporal variability. In the \chandra observation, the spectrum is best fit with an absorbed cool disc blackbody model ($T_{in}$=0.13\,keV), probably indicating the presence of a massive (or IMBH) black hole (although no short-term variability was detected). The \xmm data show similar results, but the spectra are not adequately fitted with {\it any} single component model (the best-fit single component model is a BB with $kT$=0.14\,keV, $\chi^2_\nu=1.49$). The best fit two-component model is an absorbed PL+BB model with an extremely hard PL slope and cool BB temperature ($\Gamma$=0.4, $kT$=0.12\,keV), but a PL+DISKBB model gives almost identical fit statistics with similar parameters. A fit with a PL+MEKAL component also gives an improved fit over the one-component model, and a PL+(2$\times$MEKAL model ($\Gamma$=0.3, $kT$=0.3/1.0\,keV) gives a very good fit ($\chi^2_\nu=0.96$), but these data could not constrain the temperature of the second MEKAL component. There is some indication of short-term variability in the \xmm observation, and the long-term lightcurve shows that it is also varying over longer timescales (factor of $\sim$5), demonstrating that this source is likely to be dominated by an XRB.  The unabsorbed luminosity in this observation is $L_X=1.0\times10^{39} \ergsec$, putting this source into the ULX regime. We also speculate that although partially unconstrained, the two temperature MEKAL thermal plasma components could well be emission from the superbubble surrounding the XRB.

\subsection{XMM-14}
\label{sec:xmm14}

This is another of the HNR candidates of \citet{wang99}, based on its close proximity to the optical position of SNR MF57 (4.0 arcseconds offset). Its X-ray properties in the \rosat observation were similar to those of XMM-9 (H29) with a high X-ray luminosity and lack of variability. But again, the accurate \chandra position of this source showed a clear offset of 4.8 arcseconds from the SNR position and is therefore discounted as a HNR candidate \citep{snowden01}. The \chandra data for this source are best fit with a hot disc blackbody model ($T_{in}=1.55$ keV) with an unabsorbed X-ray luminosity of $L_X=6.7\times10^{38} \ergsec$, again in the range of temperatures measured in the \asca study.  

The \xmm data for this source are best fit with absorbed PL ($\Gamma=1.86$) with an unabsorbed luminosity of $L_X=8.3\times10^{38}$ ergs s$^{-1}$, although a disc blackbody model also provides a good fit with a temperature consistent with the \chandra data ($T_{in}=1.49$ keV).  Although not shown in Table~\ref{table:spectra1}, it is also well fit by an absorbed MEKAL model  ($kT=6.4$\,keV, $\chi^2_\nu=0.49$), though this temperature is unrealistically high. The K-S test indicates possible short-term variability, and the long-term data show that it is definitely varying over long timescales (factor of $\sim$3). We therefore conclude that this is another XRB.

\section{Discussion}
\label{sec:discuss}

\subsection{Overview of Source Properties}
\label{sec:overview}

\begin{figure}
\centering
\scalebox{0.7}{\rotatebox{270}{\includegraphics{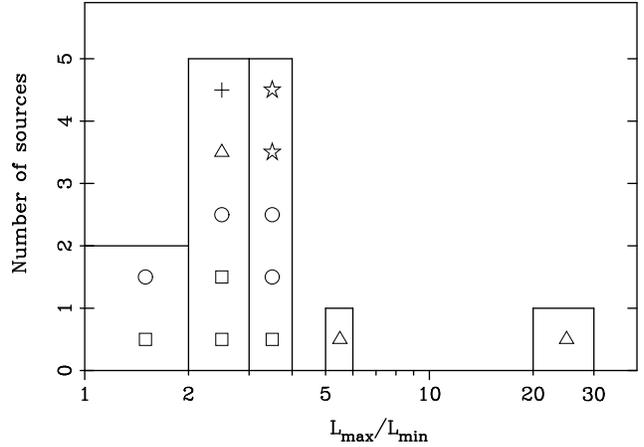}}}
\caption{Plot of the ratio between the highest and lowest observed luminosities ($L_{max}/L_{min}$) in the 0.5--2\,keV band. The symbols denote PL (squares), DISKBB (circles), supersoft (cross), PL+DISKBB/BBODY (triangles) and PL+MEKAL (stars) sources.}
\label{fig:lumratio}
\end{figure}

We have performed spectral and temporal analyses on the fourteen most luminous sources in M101 at the time of the \xmmn observation. The most striking result of the spectral analysis is that the sources appear to be a heterogeneous population. Five sources are best described by a simple absorbed DISKBB model with temperatures ranging from 0.16--1.33\,keV, three of which have unabsorbed luminosities in the ULX regime. Another four are well described by absorbed PL models with photon indices ranging from 1.80--2.55. The PL spectral fits of the remaining five sources are improved by the addition of a second soft thermal component. On the other hand, eleven of the sources show some short-term temporal variability, and the majority show long-term variability/transient behaviour.  These properties are consistent with most (if not all) of the sources in this sample being XRBs.

We investigate whether any trends in behaviour with spectral type are present in the following manner. We quantify the long-term variability by plotting the ratio between the maximum and minimum observed luminosities in the long-term data, indicating the different spectral types with different symbols (Figure ~\ref{fig:lumratio}). In each case, we have used the {\it observed} minimum and maximum values except for XMM-14, where the upper limit of the third \rosat HRI data point is the lowest value and hence the emission must be less than this. While two sources (XMM-9 \& 12) only show minor flux variability (factor of $\sim$1--2), it is clear that the majority of sources vary between a factor of $\sim$2--4, with no distinct grouping of sources with certain spectral types. It is however interesting to note that the two sources with the highest degree of long-term variability (the transient XMM-2 and XMM-13) are those with PL+DISKBB/BBODY composite spectra.

\begin{figure*}
\centering
\scalebox{0.71}{\rotatebox{270}{\includegraphics{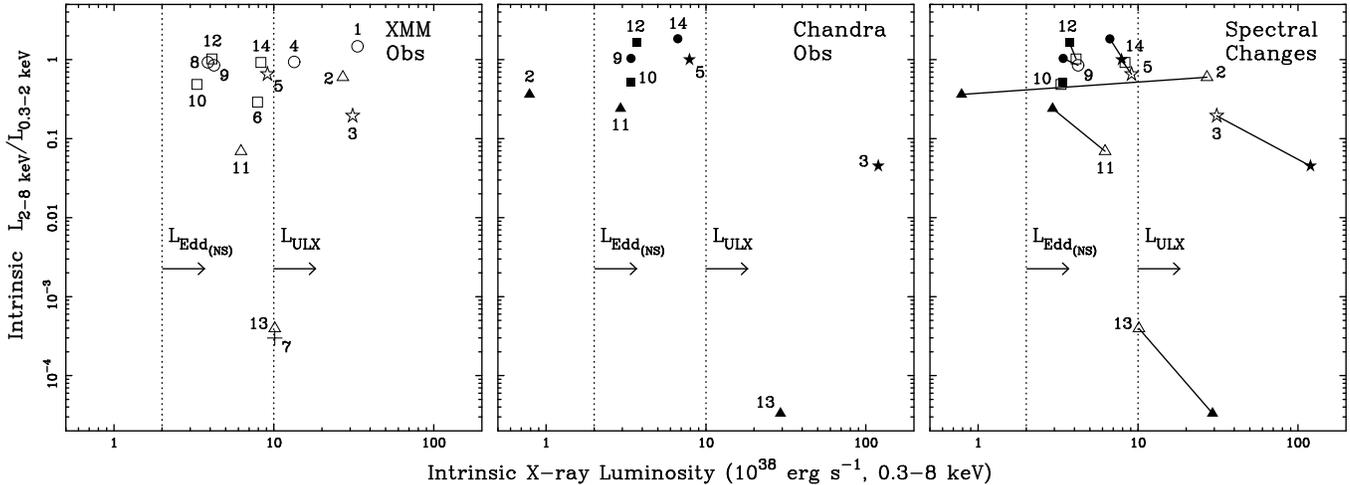}}}
\caption{Plot of the ratio between the hard ($L_{2-8 keV}$) and soft ($L_{0.5-2 keV}$) intrinsic (unabsorbed) luminosities vs the broad-band (0.3--8\,keV) intrinsic luminosity for both the \xmm [left] (open symbols) and the long \chandra [centre] (filled symbols) observations. [Right] The spectral changes are shown for those sources with spectral fits in both observations. The symbols denote PL (squares), DISKBB (circles), supersoft (cross), PL+DISKBB/BBODY (triangles) and PL+MEKAL (stars) sources.}
\label{fig:lumhard}
\end{figure*}

We can make a simple comparison in spectral shapes between the single component model fits for the long \chandra observation and the \xmmn observation of M101 (Tables~\ref{table:spectra1} \& \ref{table:spectra3}). The nine sources with sufficient counts for spectral fitting in both observations are all best fit with the same (PL or DISKBB) continuum model in both observations, except XMM-14 where both models are statistically acceptable due to the low numbers of counts.  Also, the majority of these sources show no change in powerlaw slope or disc temperature within the 90 percent confidence errors. The only significant change in shape in the whole sample is in XMM-3; the powerlaw slope hardens between \chandra and \xmm observations (3.60 to 2.78) with only a marginal decrease in flux, but  this may be offset by the difference in best-fit column densities of the two models. To illustrate the spectral hardness of these sources, in Figure~\ref{fig:lumhard} we have plotted the ratio between the hard ($L_{2-8 keV}$) and soft ($L_{0.3-2 keV}$) intrinsic (unabsorbed) luminosities versus the broad-band (0.3--8\,keV) intrinsic luminosity for both \xmm [left] and the long \chandra [centre] observations measured directly from the spectral fits.  Figure~\ref{fig:lumhard} [right] illustrates the changes in hardness and luminosity between sources with spectral fits in both observations. The majority of sources show a softening with increasing luminosity typical of Galactic black hole XRBs (BHXBs) such as Cyg X-1 (e.g. \citealt{mcclintock03}). Similar behaviour has also been observed in some ULXs (e.g. \citealt{kubota01a}; \citealt{laparola01}), though other ULXs are known to display the opposite behaviour (e.g. \citealt{fabbiano03a}; \citealt{roberts03b}). The exception to this trend in M101 is the transient source XMM-2, which shows little change in hardness despite a substantial change in overall luminosity. 

One interesting result from inspecting Figure~\ref{fig:lumhard}, and in particular the left panel, is that there does not appear to be any change in either the underlying spectral shape or in the spectral hardness across the $L_X = 10^{39} \ergsec$ ULX threshold.  Albeit on the basis of a small sample, this suggests that, in the absence of a luminosity measurement, ULXs are spectrally indistinguishable from somewhat less-luminous sources in nearby galaxies.  This in turn suggests that sources both above and immediately below the ULX threshold are intrinsically very similar, if not the same, and that no new source class is appearing in M101 above $L_X = 10^{39} \ergsec$.  Indeed, if we make the assumption that our sources are spherically accreting BHXBs at the Eddington limit ($L_E=1.5\times10^{38}(M/M_{\odot}) \ergsec$, e.g. \citealt{makishima00}), their intrinsic X-ray luminosities in the \xmm observation imply black hole masses in the range of $\sim$2--23$M_{\odot}$\footnote{Of course, the inferred luminosity of XMM-3 in the \chandra observation implies a much higher black hole mass, though a large uncertainty is associated with this due to the large absorption correction used (see Section~\ref{sec:xmm3}).}.  This is reasonably consistent with the range of measured masses for Galactic stellar-mass black holes (3--18$M_{\odot}$, \citealt{mcclintock03}), implying that there is no strong requirement for IMBHs to explain any of the present observations.  It is therefore very likely that the luminous point sources we see in M101 are simply the extreme end of the ``ordinary'' XRB population in the galaxy.  This is consistent with the cumulative luminosity functions of compact sources in other galaxies with high star formation rates, which extend to ULX luminosities and show no break at $\sim 10^{39} \ergsec$ \citep*{grimm03}.

The locations of these sources can provide clues to the nature of the X-ray emission.  However, they can also be misleading if any of the sources are background AGN shining through M101. We estimate the probability of any of our detections being background sources using the hard band (2--10\,keV) integral log($N$)--log($S$) relationship of \citet{campana01} (we judge this to be a better indicator than the soft band log($N$)--log($S$) due to the high, spatially variable neutral absorption column associated with the disc of M101). After converting our source detection threshold of 300 counts in the 0.3--10\,keV band into a 2--10\,keV flux for a typical AGN, and folding this through both the exposure map for the observation (to correct for sensitivity changes across the EPIC field-of-view) and the 2--10\,keV log($N$)--log($S$), we predict a total contamination of $\sim$2 background AGN out of fourteen sources.  Hence the likely contamination of the luminous X-ray source population with background AGN is small.

So are any of the fourteen sources good candidates to be a background AGN? The majority of the sources in this sample are actually firmly associated with star forming \hii regions or are located within spiral arms, arguing that they are intrinsic to M101. Interestingly, the outlying sources that are arguably the best candidate AGN (XMM-3, 5 \& 8) are also all located in the direction of faint outer arms of the galaxy. The relatively high neutral hydrogen columns we measure in the spectral fits are consistent (within the 90 percent confidence limits) with most of the sources being embedded within the neutral absorbing material of M101, which ranges in column between $\sim 1\times10^{20}$ cm$^{-2}$ in the interarm disc regions up to $\sim 10^{22}$ cm$^{-2}$ in the spiral arms \citep{braun95}.\footnote{This also implies that in most cases there is no strong requirement for additional absorbing material in the vicinity of the sources above that through the disc of M101.}  One final test of whether a source is a good candidate for a background AGN is if it has a bright, point-like optical counterpart.  This is the case for the source XMM-5, though its X-ray characteristics may argue against an AGN interpretation (c.f. Section~\ref{sec:xmm5}).  It is therefore difficult to identify any of the fourteen sources as an obvious background AGN.

\subsection{The nature of the compact object}

If all of our fourteen sources are accreting XRB systems, what is the nature of the compact object? Characteristics of Galactic BHXBs such as state transitions and rapid temporal variability have also been found in systems known to contain a neutron star primary. Even though all the sources have intrinsic luminosities above the Eddington limit for accretion on to a 1.4$M_{\odot}$ neutron star ($L_E^{NS}=2\times10^{38} \ergsec$), we know that both neutron star XRB (NSXB) and BHXB systems can achieve super-Eddington luminosities. For example, the three Galactic BHXB sources (V4641 Sgr, 4U 1543-47 \& GRS 1915+105) have been observed with peak luminosities well into the ULX regime \citep{mcclintock03}. Super-Eddington luminosities have also been observed for a few known NSXBs, the clearest example of which is the LMC transient X0535--668 (A0538--66) with a peak luminosity of $L_X\sim1.2\times10^{39} \ergsec$ \citep{bradt83}.

A definitive signature of a NSXB is a detection of a Type I X-ray burst, a result of the accumulation of matter onto the surface of a neutron star where it undergoes a thermonuclear flash.  These are not seen in black hole systems due to the lack of a physical surface for matter to accumulate onto (e.g. \citealt{narayan02}).  The duration of these flashes range from seconds to minutes, with intervals on time scales of hours or days. Their X-ray spectral shape is well described by a BB spectrum which softens as it decays as a result of the cooling of the neutron star photosphere. However, although the peak of the emission in strong bursts can be up to $10^3$ times higher than the persistent flux, it cannot exceed the Eddington limit \citep*{lewin95}. Most importantly, NSXBs accreting at very high rates ($L_{acc}\ga0.25L_{Edd}$) are expected to be in a stable regime where no thermonuclear bursts are seen \citep{narayan03}, and hence the absence of bursts in our source lightcurves is expected and does not assist in distinguishing between neutron star and black hole primaries.

The X-ray spectral shapes of known NSXBs and BHXBs may provide clues to the type of compact object present in these systems in M101. The observed spectra of bright NSXBs ($L_X\sim10^{38} \ergsec$) consist of two components: BB emission thought to arise in the neutron star envelope with $kT\sim2$\,keV \citep{mitsuda84}, and a softer DISKBB component from the accretion disc ($T_{in}\sim$1.5\,keV) that cools with decreasing $L_X$ \citep{tanaka96}.  When $L_X<10^{37} \ergsec$, the spectra become PL dominated. For example, the LMC transient A0538--66 displayed a BB spectrum with $kT\sim2.4$\,keV during an \einstein observation of the source in outburst \citep*{ponman84}. Alternatively, if the compact object is a black hole, the spectral shape is characterized by a soft DISKBB component with lower typical temperatures ($\leq$1.2\,keV) plus a hard powerlaw tail. Since all of our sources have $L_X>10^{38} \ergsec$, we would expect any spectra of NSXBs to have both BB+DISKBB components, but this is {\it not} the case. The only single-component source that is well fit with a BB model is XMM-7, but this displays a soft spectrum with $kT \sim$0.14\,keV, far too cool for a neutron star envelope. To investigate whether the spectra of any of the two-component sources are consistent with the compact object being a neutron star, we have attempted to fit the data with a model where the cool component is represented by a DISKBB and the hard component by a BB. This generally gave reasonable fits ($\chi^2_\nu\sim$0.9--1.41), but resulted in DISKBB temperatures of $T_{in}\sim$0.2--0.4\,keV and BB temperatures of $kT \sim$0.1--1.5\,keV, both too low to be consistent with neutron stars accreting at the Eddington limit. The only exception to this was XMM-13, where the fit yielded an unrealistically high BB temperature of $kT \sim$4.3\,keV. Hence the lack of these spectral signatures in any of the fourteen sources suggests that these systems are more likely to contain black holes rather than neutron stars.

\subsection{Single Component Spectral Fit Sources}

\subsubsection{High Temperature Disc Black Body Sources}
\label{sec:diskbb}

Four of the five \xmm sources that are best fit with a simple absorbed DISKBB model have hot inner disc temperatures ($T_{in}$=0.96--1.33\,keV), two above and two below the ULX luminosity boundary. It is now clear from RXTE studies is that the temperatures of these sources are clearly within the range observed from Galactic BHXBs in the HS (disc-dominated) emission state, which lie in the range 0.7--2.2\,keV \citep{mcclintock03}.  This again suggests a similarity between our sources and the known Galactic BHXB population.

\subsubsection{Supersoft Sources}
\label{sec:ssoft}

XMM-7 has a very soft blackbody spectrum with a temperature of $T_{in}$=0.14$^{+0.03}_{-0.03}$\,keV (BBODY) / $T_{in}$=0.16$^{+0.04}_{-0.04}$\,keV (DISKBB). Such supersoft sources (SSSs) have been detected in the Milky Way (e.g. \citealt{greiner00}) as well as a number of other galaxies (M101, \citealt{pence01}; M81, \citealt{swartz02}; M31, \citealt{distefanoetal03}; The Antennae, \citealt{fabbiano03b}; M83, \citealt{soria03}) with luminosities of $10^{37}-10^{39} \ergsec$. Although SSSs were classically defined as sources with emission dominantly below 0.5\,keV \citep{greiner00}, this has more recently been expanded to include sources with either blackbody type spectra with $kT<175$\,eV or powerlaw spectra with $\Gamma\geq3.5$, with less than 10 percent of the total luminosity above 1.5\,keV \citep{distefano03b}. This class of X-ray source may include several object types, including SNRs and accreting neutron stars. However, it is generally believed that most are a result of nuclear burning on the surface of white dwarfs (WDs) accreting at high rates, with a maximum temperature of $\sim$150\,eV corresponding to the Eddington luminosity for nuclear burning on a 1.4$M_{\odot}$ WD \citep{distefano03b}. It is therefore possible that XMM-7 with $L_X\sim10^{39} \ergsec$ is an extreme example of a WD SSS, but another interesting possibility is that the emission in extremely luminous SSSs such as this arises in the accretion discs of IMBHs, since such high luminosities coupled with such low disc temperatures imply black hole masses of $\geq100M_{\odot}$.

However, \citet{kingpounds03} show that ultrasoft components in super-Eddington sources may result from Compton-thick outflowing material from stellar-mass black holes, with effective photospheric sizes of a few Schwarzchild radius $R_S$, comparable to that expected in IMBHs. \citet{mukai03} use this outflow scenario to explain the soft X-ray emission in the M101 transient source P98 (not detected in the \xmm observation, see section~\ref{sec:trans}). \citet{fabbiano03b} discuss a highly variable supersoft ULX in the Antennae, and demonstrate that its changing spectral characteristics can be explained a 10$M_{\odot}$ black hole with a mass outflow rate consistent with that expected from thermal timescale mass transfer in a massive XRB or for a bright soft X-ray transient (SXT) burst, the two scenarios put forward to explain ULXs by \citet{king02a}. 

Although we do not have sufficient data to test whether the spectral shape of XMM-7 varies with time, its 0.5--2\,keV emission does show some variability on long timescales, which argues against the alternative interpretation for this source of a HNR (see section~\ref{sec:xmm7}). It has an intrinsic luminosity $\sim10^{39} \ergsec$ in the \xmm observation and may indeed be a stellar-mass XRB with an outflow. Future observations of this source with \xmmn or \chandra may give additional clues to its nature.

\subsubsection{Powerlaw Sources ?}
\label{sec:PL}

Another result of this analysis is that the spectra of four sources (XMM-6, 10, 12 \& 14) are best-fit with simple absorbed powerlaw continua with $\Gamma\sim$1.8--2.6. Luminous sources with powerlaw spectra (as opposed to DISKBB spectra) have been found in recent studies e.g. the \xmmn minisurvey of ULXs \citep{foschini02}; NGC~5204 X-1 \citep{roberts01}; M51 (NGC~5194 \#37 \& 82, \citealt{terashima03}); M83 \citep{soria03}; NGC~4485/90 \citep{roberts02}.  While powerlaw spectra with $\Gamma$=1.5--2 are typical of the low/hard (LH) state of Galactic BHXBs \citep{tanaka96}, steeper spectral shapes could represent strong Comptonized thermal emission from a black hole accretion disc in a very high/steep powerlaw state (VHS/SPL). Based on {\it RXTE} observations, \citet{mcclintock03} define the SPL state by the presence of a powerlaw component with $\Gamma>2.4$. Out of these four \xmm sources, only XMM-6 \& 10 fall within this category (within the 90 percent confidence limits), while XMM-12 \& 14 have harder powerlaw slopes of $\Gamma\sim1.8-1.9$ consistent with the LH state.

However, in this dataset we {\it cannot} reject the DISKBB model fits for three of four of these sources (XMM-6, 12 \& 14). Although these are marginally better fit with the PL model (see comparison of ratios in Figure~\ref{fig:spec1}), the DISKBB fits for these sources all give $\chi^2_\nu\la1$ with disc temperatures in the range of 0.7--1.5\,keV, similar to the sources discussed in section~\ref{sec:diskbb}. {\it It is therefore entirely possible that eight out of nine single component sources are XRBs in the HS state}. We explore this idea further by investigating whether the underlying hard continuum in the multi-component sources can be modelled with a DISKBB in section~\ref{sec:2comp}.

An extremely interesting result is that only one source, the nuclear source XMM-10, has an unambiguous powerlaw spectral shape with $\Gamma=2.3$. This marks XMM-10 as an unique source in the sample, and could be interpreted as evidence that the nucleus of M101 harbours a low-luminosity AGN (LLAGN), although the spectrum is steeper than the standard AGN photon index ($\Gamma\sim$1.7--2.1, \citealt{nandra94}). XMM-10 is resolved into two sources in the \chandra data, both with powerlaw spectra (see section~\ref{sec:xmm10}), with the source coinciding with the optical nucleus (within $\sim$35\,pc) possessing the steepest spectrum of $\Gamma=2.1$. The lightcurve data for XMM-10 shows that at least one of the two \chandra sources is variable on both long- and short-timescales, but it is impossible to distinguish which of the two may be dominating the emission in the \xmm data.

As a comparison, \citet{baganoff03} have recently reported that the \chandra spectrum of the X-ray emission from the Sagittarius A$^{\ast}$ massive dark object at the centre of our own Galaxy (morphologically similar to M101) can be fitted with a steep powerlaw ($\Gamma=2.7$) spectrum, albeit with an extremely low luminosity of $2.4\times10^{33} \ergsec$ (0.5--7\,keV), consistent with a quiescent super-massive black hole (SMBH). This is a factor $\sim10^5$ less luminous than XMM-10.  Interestingly, there is no evidence for an AGN in M101 in the optical regime, with the nucleus classified as an \hii region by \citet*{ho97} based on its emission-line ratios. It is therefore perfectly plausible that we are simply detecting a bright XRB in the vicinity of the nucleus.

\subsection{Multi-component Sources}
\label{sec:2comp}

Five of our luminous sources are best fit with multi-component spectral models. Similarly, two-component models have been required to adequately fit some other recent \xmm and \chandra data e.g. NGC~1313 X-1 and X-2 \citep{miller03a}; NGC~6946 X-11 \citep{roberts03a}; two M51 ULXs \citep{terashima03}; NGC~5408 X-1 \citep{kaaret03}. Even though our most luminous source (XMM-1) is well fit with a simple DISKBB model, the other sources provide evidence that many sources could have more complicated spectral shapes than are shown by the simple fits that can be done on photon-limited data.

To begin with, we fitted these sources with the canonical model of a powerlaw continuum plus a soft component to model the soft excess. Three are best fit with PL+DISKBB/BB models (XMM-2, 11 \& 13), although there is little difference statistically between the DISKBB and BB fits. The disc temperatures are cool, and in two cases (XMM-11 \& 13) are similar to the inner disc temperatures of SSSs discussed in section~\ref{sec:ssoft}. \citet{miller03a} find similar disc temperatures in their study of two ULX sources in NGC~1313, and interpret this as evidence of IMBHs based on the argument that the black hole mass scales inversely with the accretion disc temperature ($T_{in}\propto M^{-1/4}$). The ambiguity of which type of blackbody model is the best fit plus the fact that our source luminosities do not imply IMBH masses casts some doubt on this interpretation in our data. Interestingly the powerlaw slopes are all $<2$ (XMM-13 has a particularly hard slope with $\Gamma$=0.36), harder than seen in the HS state of Galactic BHXB candidates ($\Gamma$=2.1--4.8, \citealt{mcclintock03}), and all have a substantial contribution to the total source fluxes.

The other two sources (XMM-3 \& 5), both with ULX luminosities, are best fit with PL+MEKAL models. As discussed in sections~\ref{sec:xmm3} \& \ref{sec:xmm5}, both are located far from the central region of the galaxy where most of the diffuse thermal gas is found, and so it is unlikely that what we detect here is contamination from the diffuse emission in the disc of M101. We are therefore likely to be detecting thermal plasma in the vicinity of the hard X-ray source. Although this could suggest the presence of SNRs, none was detected at the location of XMM-3 by \citet{matonick97}, though the position of XMM-5 is just outside the field covered in that study. In the case of XMM-3, the MEKAL temperature is hot ($kT$=0.98\,keV) and it may be the case that we are detecting hot photoionized plasma surrounding a high-mass companion star in a binary system. Galactic high-mass XRBs (HMXBs) can exihibit strong recombination lines if accreting from highly ionized stellar winds (e.g. Cyg X-3, \citealt{soria03} and references therein). \citet{terashima03} point out that the luminosity of this thermal component is likely to constitute $<10$ percent of the total flux of the system, which is consistent with the 5 percent MEKAL contribution we measure here. In the case of XMM-5, the plasma component makes a much more substantial contribution to the total source flux (20 percent), but its temperature is much lower ($kT$=0.17\,keV), and we could be seeing thermal emission related to star formation reminiscent of the low-temperature component of starburst superwinds (e.g. NGC~253, \citealt{pietsch01}, \citealt{strickland00}; NGC~3256, \citealt{lira02}), although lack of evidence for star formation activity at other wavelengths is problematic. The powerlaw components in these sources are soft ($\Gamma>2$), similar to those of XMM-6 \& 10 discussed in the previous section. XMM-13, located in the giant \hii region NGC~5461, is another interesting case, as mentioned in section~\ref{sec:xmm13}. Although not fully constrained with these data, the best fit is a two-temperature MEKAL plus hard PL model ($\Gamma$=0.3, $kT$=0.3/1.0\,keV), where the MEKAL components constitute $\sim$43 percent of the total flux of the source, again reminiscent of starburst superwind plasma temperatures which could perhaps arise in a very energetic \hii region.

Since we are speculating that all of the single-component spectral fit sources in M101 (except the nuclear source XMM-10) are accreting XRBs in a high state (see section~\ref{sec:PL}), we have also attempted to model the underlying hard continuum in the multi-component sources with a DISKBB with temperatures similar to those found in stellar-mass black holes ($\sim$1--2\,keV). In three cases (XMM-3, 5 \& 11), this was successful, with statistical improvements at the $\ga$ 99 percent level compared with the single-component DISKBB fits.  XMM-3 could be modelled by both DISKBB+PL ($T_{in}$=0.9\,keV, $\Gamma$=3.2, $\chi_{\nu}^2\sim1.1$) and DISKBB+BBODY ($T_{in}$=0.9\,keV, $kT$=0.2\,keV, $\chi_{\nu}^2\sim1.0$) models; XMM-5 with a DISKBB+MEKAL model ($T_{in}$=1.2, $kT$=0.2\,keV, $\chi_{\nu}^2\sim1.4$) and XMM-11 with a DISKBB+BBODY model ($T_{in}$=1.5\,keV, $kT$=0.2\,keV, $\chi_{\nu}^2\sim1.3$). The fits for the remaining two sources (XMM-2 \& 13) yielded unrealistically high disc temperatures ($\ga$ 3\,keV). This interpretation of the data has important implications for the IMBH scenario. The hard continuum in such sources is usually interpreted as powerlaw emission similar to that seen in the LH state of Galactic XRBs; if the soft excess is modelled by a cool DISKBB, the best-fit temperatures imply black hole masses in the IMBH range. However, if the hard component can be equally successfully modelled by a DISKBB with a ``normal'' stellar-mass black hole temperature, then this component could represent the thermal signature of the accretion disc (with a temperature typical of a stellar-mass black hole) and the soft component must have a separate origin.  Again this implies that an IMBH is not required.  Further detailed spectral studies of similar sources are required to explore this possibility.

\subsection{Transient Sources}
\label{sec:trans}

It is believed that most low-mass XRBs (LMXBs) will exhibit transient behaviour during their lifetimes \citep{king02a}, in the form of SXT outbursts caused by hydrogen-ionization instabilities in the accretion disc around the black hole or neutron star.  The most obvious transient source we have detected in this study is XMM-2, which has been in a quiescent state in all previous observations and has brightened by a factor of $\sim$30 in the \xmm observation. It also is worth commenting on the lack of a detection in the \xmm observation, and therefore the transient behaviour, of the most luminous ULX (P98) in the \chandra observation, a highly variable source which seemed to reflect anticorrelated changes in the source temperature and size \citep{mukai03}. Assuming the model parameters of \citet{pence01}, we calculate an upper flux limit of $\sim 8\times10^{-15} \ergcms$ in the \xmm data, which implies a drop in luminosity of at least a factor of $\sim$20 since the \chandra observation. Such transient sources are likely to be LMXBs, since known black hole HMXB systems are persistent X-ray sources \citep{mcclintock03}\footnote{Note that though SXT outbursts may occur in HMXB systems with Be-star companions, these systems possess a neutron star primary \citep{tauris03}.}, although the caveat here is that the known black hole HMXB systems are wind accretors, i.e. they are not going through a phase of thermal-timescale mass transfer that may be necessary to reach ULX luminosities (see \citealt{kalogera03} for more detailed discussions). Future monitoring observations of galaxies such a M101, designed to detect transience over timescales of months, might therefore provide a means of distinguishing between LMXB and HMXBs possessing black hole primaries, which cannot be distinguished spectroscopically in the X-ray regime.

\section{Conclusions}
\label{sec:summary}

In this paper, we have studied the X-ray spectral and temporal properties of the most luminous sources in M101 at the time of the \xmmn observation, and our results suggest that they are all accreting systems associated with M101. For comparison purposes, we have also analysed two archival \chandra observations of M101 to investigate spectral and temporal changes in these sources. The positions of thirteen of the fourteen \xmm sources are covered by either or both of the \chandra observations. Of these, eleven are unresolved by {\it Chandra}, while the remaining two (XMM-2 and the nuclear source XMM-10) are both resolved into two discrete sources. Our results can be summarized as follows:

\begin{itemize}

\item Fourteen sources in the \xmm observation have sufficient counts for spectral fitting, with intrinsic luminosities in the range of 3.4$\times10^{38}$--3.4$\times10^{39} \ergsec$. If these are BHXB systems accreting at the Eddington limit, these luminosities imply black hole masses of $\sim2-23M_{\odot}$, consistent with the measured range of Galactic stellar-mass black holes.

\item The locations of the sources suggest that they are all associated with M101 (\hii regions/spiral arm/nuclear), and suggest a link between the high-luminosity XRB population and young stellar populations.

\item Five sources are best described by simple DISKBB models. Four have hot inner disc temperatures ($T_{in}$=0.96--1.33\,keV), consistent with the disc temperatures found in Galactic BHXBs ($T_{in}$=0.7--2.2\,keV, \citealt{mcclintock03}) in the HS state. 

\item Four sources have simple absorbed powerlaw spectra with $\Gamma\sim$1.8--2.6, two of which are consistent with the SPL state. However, three of these sources are also well modelled by DISKBB models with disc temperatures of 0.7--1.5\,keV, making it {\it entirely possible that eight out of nine sources single-component sources are HS XRBs}. Only the nuclear source (XMM-10) has an unambiguous powerlaw spectrum, which may be evidence of either an LLAGN or alternatively an XRB (given no indication of AGN activity at other wavelengths).

\item Five sources require multi-component spectral fits, in the form of an underlying hard PL continuum plus a soft excess. In three sources, the soft component can be modelled by either DISKBB/BBODY models, two of which have cool disc temperatures which could be interpreted either as evidence for the presence of IMBHs or as the signature of a Compton-thick outflow from a stellar-mass black hole. In two (possibly three) sources, the soft excess is well modelled with one or more MEKAL thermal plasma components. The hot ($kT$=0.98\,keV) component in XMM-3 could arise in a hot photoionised plasma surrounding a high-mass companion star, and the cooler ($kT$=0.17\,keV) component detected in XMM-5 could be thermal emission related to star formation activity with a temperature similar to those of starburst superwinds. If we accept the PL+(2xMEKAL) interpretation of the data for XMM-13, we could be detecting a multi-temperature plasma from the superbubble encompassing an XRB. In three cases (XMM-3, 5 \& 11), the underlying hard component can also be successfully modelled by a hot DISKBB with a ``normal'' stellar-mass black hole accretion disc temperature ($T_{in}$=0.9--1.5), arguing against the IMBH scenario.  

\item Eleven sources show statistical evidence of short-term variability during the \xmm observation. The long-term data demonstrate that most are also variable over a baseline of 11--24 years, with the majority varying between a factor of $\sim$2--4 in observed luminosity. This variability is strong evidence in favour or these sources being accreting XRB systems.

\item In the absence of the spectral signatures expected from neutron stars in outburst, we conclude that the accreting objects in these binary systems are more likely to be stellar-mass black holes.

\item We find no evidence to support the HNR scenario in those sources which have previously been suggested to be. Evidence of long- and short-term variability coupled with the DISKBB spectral shape of XMM-1 effectively rules out the HNR scenario for this source. Although we find no conclusive evidence to rule out XMM-7 (NGC~5471B) as a HNR, it has a supersoft \xmm spectrum, which may result from nuclear burning on the surface of a WD accreting at a high rate, a Compton-thick outflow of material from a stellar-mass black hole or the accretion disc of an IMBH.

\item Even though the sources show a variety of spectral shapes and hardness, there is {\it no apparent spectral distinction between those above and below the ULX luminosity threshold} ($10^{39} \ergsec$), implying that we are seeing the extreme high-luminosity end of a ``normal'' XRB population.

\item Out of the nine sources with spectral fits in both the \xmm and \chandra observations, no major spectral changes are apparent. While the majority do show some degree of spectral softening with increasing luminosity as seen in Galactic BHXBs, the transient source XMM-2 shows little change of hardness despite a significant change in luminosity.

\item Two transient sources have been detected in these observations; XMM-2 in the \xmm observation and the \chandra source P98 studied by \citet{mukai03} which then fell below the detection threshold of the \xmm data. We speculate that since LMXBs are known to exhibit transient behaviour, future multiple observations of galaxies such as M101 may provide a way to distinguish between LMXBs and HMXBs.

\end{itemize}

\vspace{3mm}

\noindent We have now reached the era of X-ray astronomy in which detailed studies of the most luminous individual XRBs beyond the Local Group (SMC, LMC, M31) are possible. This paper demonstrates that these sources, the extreme examples of accreting stellar remnants, display a wide range of X-ray characteristics. With the superb capabilities of \xmmn and {\it Chandra}, we now have an exciting opportunity to begin to expand our knowledge of these and other objects apparently accreting at or above the Eddington Limit.

\section*{Acknowledgments}

We thank the anonymous referee for very helpful comments. This work is based on observations obtained with {\it XMM-Newton}, an ESA science mission with instruments and contributions directly funded be ESA and NASA. The second digitized sky survey was produced by the Space Telescope Science Institute, under Contract No. NAS 5-26555 with the National Aeronautics and Space Administration.  LPJ gratefully acknowledges the support of a PPARC studentship.

\label{lastpage}

{}

\end{document}